\documentclass[aps,prd,nofootinbib]{revtex4}
\usepackage{graphicx}
\def\s#1{\slash\!\!\!{#1}}
\begin{document}
\baselineskip=20pt
\title{Some aspects of neutrino astrophysics\footnote{
 Based on lectures given at The Sixth Constantine High Energy Physics School on 
 {\em Strong
 and Weak Interactions Phenomenology}, 6 $-$ 12 April, 2002, Constantine, Algeria.}}
\author{H. Athar}
\affiliation{Physics Division, National Center for Theoretical
         Sciences, 101 Section 2, Kuang Fu Road,
         Hsinchu 300, Taiwan,
         E-mail: athar@phys.cts.nthu.edu.tw}
\date{\today}
\begin{abstract}

Selected topics in  neutrino astrophysics are reviewed. These
include  the production of low energy neutrino flux from cores of
collapsing stars and the expected high energy neutrino flux from some
other astrophysical sites such as the galactic plane as well as
the center of some distant galaxies. The expected changes in these
neutrino fluxes because of neutrino oscillations during their
propagation to us are described. Observational signatures for these
neutrino fluxes with and without neutrino oscillations are
discussed.

\end{abstract}
\maketitle
\section{General Introduction}
    In the standard model of the electro weak interactions, the
lepton masses and the values of other parameters such as weak
mixing angle, couplings, etc. are arbitrary and are therefore
determined by experiments.
These parameters are independent of each other and can not be
determined uniquely, while the neutrino is taken to be massless
because of maximal parity violation. The masslessness of the
neutrino however does not follow from any other theoretical ground unlike
the local gauge invariance for the photon \cite{Fukugita:wx,Kim:dy,djouadi}.

    In order to search for physics {\tt beyond} the standard model of particle 
 physics, light
neutrino masses are incorporated in the extensions of the
standard model. Then, on quite general grounds one expects that
neutrinos will also possess non-zero magnetic moments.
 Experimentally, one places finite non-vanishing upper bounds
 on measurable neutrino masses and
    magnetic moments. Presently, there is some indirect
    experimental evidence for the masslessness of the neutrino \cite{Valle:2002tm}.

     Massive neutrinos quite likely mix.
    The mixing of quarks is an established fact and because of
    quark-lepton symmetry, it is natural to assume that leptons
    exhibit mixing as well. An additional argument, in this sense,
    is provided by grand unification models, in which quarks and
    leptons are described in a unified manner.

    Mixing means that $\nu_{e}$, $\nu_{\mu}$ and $\nu_{\tau}$,
i.e., the states created in weak interactions, are different from
the states $\nu_{1}$, $\nu_{2}$ and $\nu_{3}$ that have definite
masses. The neutrinos $\nu_{e}$, $\nu_{\mu}$ and $\nu_{\tau}$ are
orthogonal combinations of $\nu_{1}$, $\nu_{2}$ and $\nu_{3}$ with
different phases between them. In addition, the sterile neutrinos
may mix with these neutrinos.

    Neutrino mixing has, as its consequence, neutrino
    oscillations, i.e., the process of periodic (complete or
    partial) conversion of neutrinos of one type into another, for
    instance, $\nu_{e}\to \nu_{\mu}\to \nu_{e}\to ... $ . The
    components $\nu_{i}$ of a mixed neutrino have different masses
    and hence different phase velocities. It follows that the
    phase difference caused by the mass difference between the
    $\nu_{i}$ vary monotonically during the propagation. This
    phase change manifests itself as neutrino oscillations.

If neutrino oscillations do occur in vacuum, matter can enhance
their depth (probability amplitude) up to a maximal value
\cite{Mikheev:qk}. That is, a monotonic change of density may lead
to resonant conversions between various neutrino flavors. This
follows from the fact that when neutrinos propagate through a
monotonically changing density medium, $\nu_{e}$ and $\nu_{\mu}$
($\nu_{\tau}$) feel different potentials, because $\nu_{e}$
scatters off electrons via both neutral and charged currents,
whereas $\nu_{\mu}(\nu_{\tau})$ scatters off electrons only via
the neutral current. This induces a coherent effect in which
maximal conversion of $\nu_{e}$ into $\nu_{\mu}$ take place (even
for a rather small intrinsic mixing angle in the vacuum), when the
phase difference arising from the potential difference between the
two neutrinos cancel the phase caused by the mass difference in
the vacuum \cite{Wolfenstein:1977ue}.

 During nearly past half a century, the empirical search for neutrinos has
spanned roughly four orders of magnitude in neutrino energy $E$, from $\sim $ MeV
up to $\sim  10^{6}$ MeV. The lower energy edge corresponds to the 
Solar neutrinos, whereas the upper energy edge corresponds to the
Atmospheric neutrinos. A detailed early description of the Solar
neutrino search can be found in \cite{Bahcall:gw}, whereas for recent status, see
\cite{lipari}. The aspects of
neutrino production in Atmosphere of earth related to neutrino oscillation
studies are recently reviewed in \cite{Gaisser:2002jj}. The intermediate energy range
corresponds to terrestrial neutrinos such as from reactors (and accelerators) and the  
 supernova neutrinos. Thus, obviously either going in
energy range below these values or {\tt above} are the available
frontiers.   For a general introduction of the
possibility of having neutrinos with energy $> 10^{6}$
MeV, see \cite{Bahcall:iz}, the upper energy edge for these high 
 energy neutrinos is limited only by the concerned experiments. 
 More detailed general discussions in the context of high energy 
 neutrinos can be found in  \cite{Ginzburg:sk,Learned:sw}.
 Despite the availability of the
somewhat detailed discussion of progress cited in the last reference, the field of high 
energy neutrino astrophysics is still passing through its 
initial stage of development.

The above mentioned empirical search has already given us quite
useful insight into neutrino intrinsic properties such as
mass and mixing. 
Massive neutrinos and their associated properties such as Dirac or
Majorana character of their mass, their mixings and magnetic
moments can have important consequences in astrophysics. In these
two lectures, I elaborate some of the selected  consequences and the
constraints implied by these consequences on neutrino properties
as well as the insight that one may gain about the nature of the
astrophysical or/and cosmological sites and the interactions that 
 produce these neutrinos.
 The explanation of observed Solar $\nu_{e}$ deficit relative to its 
production value in the core of the Sun, via $\nu_{e}\to \nu_{\mu }, \nu_{\tau} $ 
conversion is an interesting example in this context \cite{Smirnov:2002in}.

The general plan of the lectures is as follows.  
 In the first lecture, I
elaborate these aspects for the low energy neutrino flux ($E \sim $
MeV) emitted during the gravitational collapse of stars which may
be accompanied by supernovae phenomenon. In the second
lecture, I elaborate these for the expected high energy neutrino
flux ($E \geq 10^{6}$ MeV) from some representative examples of 
 remaining cosmos around us such as our galactic plane. This includes the one 
 arising from the interaction of ultra
high energy cosmic ray flux ($E \geq 10^{12}$ MeV) with the matter
and radiation inside the sources (such as center of our and other
galaxies) of as well as during propagation of ultra high energy
cosmic ray flux to us.

\pagebreak

\section{Lecture 1: Neutrinos from Collapsing Stars}
\label{one}
\subsection{Introduction}
 Almost $\sim 99\%$ of the binding energy of a few Solar mass collapsing core
of a star is released in the form of neutrinos typically in an
order of magnitude energy interval between 5 and 50 MeV. In
contrast, in the Sun, the energy released from the core 
 in the form of neutrinos
is of the order of $\sim 1\%$. Therefore, neutrinos from
collapsing cores are expected to bring relatively more useful information
about the mechanism of the collapse as well as information about
the intrinsic properties of neutrinos itself, though still 
 difficulties remain in obtaining a self consistent and 
 a clear understanding of 
 the exact mechanism of supernova phenomena. The rather
optimistic frequency of occurrence of 1987 A like supernova is roughly 
once per thirty years. The last one occurred in 1987 implying that
it is by now already half way. On the other hand, the estimates of cumulative 
 neutrino flux from all the supernovae that occurred in the past 
 are also rather close to the present relevant detection threshold \cite{Malek:2002ns}.
 For a more detailed introduction, see
 \cite{Shapiro:du,Burrows:kf,Mohapatra:rq,Raffelt:2002tu}.
\subsection{Neutrino production}
Type II supernovae are the most interesting from the point of view
of neutrino flux studies. Present theories of type II supernovae
assume that they are catastrophic endpoints in the evolution of
massive stars with $(8-10)\leq M/M_{\odot}\leq 70$, where
$M_{\odot}\sim 2\cdot 10^{33}$ g is the Solar mass. 
 Such stars form unstable iron cores (or if
the mass is closer to the lower limit, O, Ne or Mg cores)
supported mainly by degenerate electron gas pressure. Partial
photo dissociation of nuclei (for instance, $\gamma
+{}^{56}_{26}Fe\to 13\alpha +4 n$) and electron capture by free
protons ($e^{-}+p\to n +\nu_{e}$) as well as by nuclei
[$e^{-}+(Z,A)\to (Z-1,A)+\nu_{e}$], known as {\tt neutronization},
causes the core to undergo a dynamic collapse when its mass
exceeds the Chandraseakhar mass $M_{\rm Ch}$ (which is function of
$Y_{e}$), namely, when $M > M_{\rm Ch}\equiv 5.83
Y^{2}_{e}M_{\odot}$. Here $Y_{e}= n_{e}/n $ is the number of
electrons per baryon. It is estimated that most of the energy
during this collapse is carried by the neutrinos. During the
collapse, the neutrinos are emitted not only by neutronization but
also by the {\tt thermal emission}. The significant thermal emission
 interactions include: pair annihilation ($e^{+}+e^{-}\to \nu
+\bar{\nu}$), plasmon decay (plasma excitation $\to \nu
+\bar{\nu}$), photo annihilation ($\gamma+ e^{-}\to
e^{-}+\nu+\bar{\nu}$) and neutrino bremsstrahlung [$e^{-}+(Z,
A)\to (Z, A)+e^{-}+\nu +\bar{\nu}$]. In all these interactions, the
electron neutrino pair can be produced by $Z$ or by the $W^{\pm}$
exchange, whereas the muon and tau neutrino pairs can be produced
only by $Z$ exchange. However, each neutrino emission interaction has
an inverse interaction corresponding to absorption. Both absorption
and inelastic interactions impede the free escape of neutrinos from
a collapsing core. The relevant important interactions include: free
interactions ($\nu +n \to \nu +n$), interactions of heavy nuclei with
$A >$ 1 [$\nu+(Z,A)\to \nu +(Z,A)$], nucleon absorption
($\nu_{e}+n\to p+e^{-}$) and electron neutrino interaction ($e^{-}+\nu
\to e^{-}+\nu$). Similar interactions occur for anti neutrinos. The
cross section for these interactions defines the depth of various
{\tt neutrino spheres}. These interactions tend to thermalize the
various neutrinos and thus contribute to neutrino opacity.

    Thus, the core of a collapsing star is the source of all
neutrino and anti neutrino species. The fluxes change with time and
are different for different species. During the initial stages of
collapse, $\nu_{e}$ flux from the neutronization dominates. On the
$\nu $ opacity stage, comparable (but not equal) fluxes of all
flavors are emitted. Due to the difference in inelastic
interactions of $\nu_{e}$ and $\nu_{\mu}$ ($\nu_{\tau}$), the
$\nu_{\mu}$ ($\nu_{\tau}$) neutrino spheres lie deeper (and
consequently at higher temperatures) than the $\nu_{e}$ neutrino
sphere. This results in the higher energies of $\nu_{\mu}$
($\nu_{\tau}$) with respect to the $\nu_{e}$ energies. Typically
$\bar{E}_{\nu_{\mu}} = (2-3)\cdot \bar{E}_{\nu_{e}}$ but in the
smaller flux of these neutrinos:
$F^{0}_{\nu_{\mu}}=\frac{1}{2}\cdot F^{0}_{\nu_{e}}$. The energies as well as
fluxes of {\tt non electron} type neutrinos ($\nu_{\mu},
\nu_{\tau}, \bar{\nu}_{\mu}, \bar{\nu}_{\tau}$) are considered to
be roughly equal. The spectra of various neutrinos being (approximately) Fermi-Dirac
spectra with different temperatures and with high energy cut.

    Summarizing, the neutrino radiation of a type II supernova
    consists of two components: a $\sim $ 10 ms electron
    neutrino burst from the neutronization of dense matter which
    is followed by $\sim $ 10$-$20 s thermal radiation of
    $\nu-\bar{\nu}$ pairs of all types; $\nu_{e}$, $\bar{\nu}_{e}$,
    $\nu_{\mu}$, $\bar{\nu}_{\mu}$, $\nu_{\tau}$,
    $\bar{\nu}_{\tau}$ typically with the above characteristics.
    The thermal $\nu$'s are emitted via black body radiation from
    the surfaces of the corresponding neutrino spheres. Their
    fluxes and spectra are therefore determined by the neutrino
    spheres, which in turn are essentially determined by the $\nu
    $ interaction cross section as mentioned earlier.

The relevant feature of the above characteristics for present
elaboration is that the neutrinos being produced in the central part of
collapsing the star and crossing the matter layers with decreasing 
 density (from $\rho 
 \sim  10^{14}$ g/cm$^{3}$ to approximately zero) may
undergo {\tt resonant conversions}. There are in general two aspects of
this phenomena. The resonant oscillations may change the
properties of $\nu $ burst which is important for the burst
detection. Moreover, the change of the $\nu $ fluxes may influence
the evolution of star: the dynamics of the collapse itself and of
the expelling envelope. I intend to elaborate the former
aspect in this lecture. The basic characteristic features of the
neutrino fluxes, i.e., the flux spectra of the various neutrino
species, etc., are the main testable ingredients of not only a
stellar collapse theory but also of a possible occurrence of
matter enhanced neutrino oscillations. These can be studied by
existing and future detectors in some detail for neutrino
bursts from nearby supernovae ($\leq 10$  kpc, where 1 pc $\sim 3\cdot 10^{18}$ cm).

\subsection{Oscillations during propagation: Effects of neutrino mixing}
Because the matter density inside a collapsing star is quite high
so that a necessary condition for resonant neutrino conversions, 
 namely the level crossing or
resonance is satisfied (see later). I, therefore, will elaborate only the effect of
 matter enhanced neutrino oscillations on the neutrino flux inside the
matter and magnetic fields of collapsing stars. There are no matter enhanced 
spin flavor oscillation effects between a nearby supernova and the earth
 for downward going neutrinos.

A large range of neutrino mixing parameters determining the (two, three and four)
flavor oscillations are presumably get measured in
terrestrial neutrino oscillation experiments with quite a good
accuracy in near future perhaps before the next nearby supernova occurrence. 
This statement is further supported when one includes the already existing 
information on neutrino mixing parameters from Solar and Atmospheric
 neutrino flux measurements. The pure neutrino flavor oscillation 
effects may thus possibly be disentangled from for instance 
 {\tt pure neutrino spin flavor oscillation} effects for supernova neutrinos. 
In the possible presence of relatively strong magnetic fields in
collapsing stars, the role of neutrino spin flavor oscillations
become relevant. Therefore, I mainly elaborate here the effects of pure
neutrino spin flavor oscillations of the type 
 $\nu_{e}\leftrightarrow \bar{\nu}_{\mu}$, $\bar{\nu}_{e}\leftrightarrow 
 \nu_{\mu}$ in collapsing stars, assuming
the smallness of neutrino flavor mixing for illustration only. 
 The neutrino spin flavor oscillations can take place for Majorana
neutrinos for vanishing vacuum mixing also as the Majorana type
neutrino magnetic moment can mix both the helicity and flavor of the
two neutrino states. A recent
detailed discussion on pure flavor oscillation effects for
neutrinos from collapsing stars is given in \cite{Dighe:1999bi}.
At the end of this subsection, I will briefly compare the observational
signatures of pure flavor and pure spin flavor oscillations
 (in absence of flavor mixing) for completeness. 
 The general case of neutrino spin flavor
oscillations in the presence of non vanishing vacuum mixing is briefly 
discussed in \cite{ahriche}.

Neutrino spin flavor oscillations are commonly studied by
numerically solving a Schrodinger like system of equations with an
effective Hamiltonian. Using the notation of \cite{Grimus:1993fz}, I briefly outline the
main steps to obtain it in a particular neutrino
basis. I  start from a most general lagrangian density, ${\cal L}$,
describing the neutrino propagation in the presence of varying
matter density and electromagnetic fields:
\begin{equation}
{\rm \cal{ L}} = \dot{\imath} \bar{\psi}_{L}\s{\partial}
 \psi_{L}+\dot{\imath} \bar{\psi}_{R} \s{\partial} \psi_{R}-
 \bar{\psi}_{L} h_{L} \psi_{L}-
 \bar{\psi}_{R} h_{R} \psi_{R}-
 \bar{\psi}_{L} h_{LR} \psi_{R}-
 \bar{\psi}_{R} h_{RL} \psi_{L},
\end{equation}
where $h_{L}$, $h_{R}$, $h_{LR}$, $h_{RL}$ are matrices with
lepton flavor and Dirac indices. These matrices are in general
space time dependent. In the Dirac case, $\psi_{L}$ and
$\psi_{R}$ are independent fields and thus ${\cal L}$ conserves
total lepton number.  The equation of motions that follow from
${\cal L}$ can be written as
\begin{equation}
\label{evolution}
 \frac{\partial \psi}{\partial t} = H_{D} \psi .
\end{equation}
Here, $\psi=\psi_{L}+\psi_{R}$ and  $H_{D}$ is given by
\begin{equation}
 H_{D}=-\dot{\imath} \underline{\alpha}\cdot
 \underline{\nabla}+ \beta
 (h_{L}P_{L}+h_{R}P_{R}+h_{LR}P_{R}+h_{RL}P_{L}).
 \end{equation}
The left and right-handedness is defined by
\begin{equation}
 P_{L}\psi_{L} = \psi_{L}, \, \, \, P_{R}\psi_{R} = \psi_{R},\, \,  {\rm with}\, \,
 P_{L,R}=\frac{1}{2}(1\mp \gamma_{5}),
\end{equation}
and $\alpha_{j}\equiv \gamma^{0}\gamma^{j}$ ($j=1,2,3$) and $\beta
\equiv \gamma^{0}$.

    A equation similar to Eq. (\ref{evolution}) can be obtained for
    Majorana neutrinos.
The total Hamiltonian comprising both is
\begin{equation}
 H_{T}=-\dot{\imath} \underline{\alpha}\cdot \underline{\nabla} +
 V_{L}P_{L}+V_{R}P_{R}+V_{S}\beta
 P_{L}+V^{\dag}_{S}\beta P_{R}+\frac{1}{2}\beta (\omega
 P_{L}+\omega^{\dag}P_{R})\sigma_{ab} F^{ab},
\end{equation}
where $V$'s are in general space time dependent potentials in
matrix form in flavor basis. These describe the effects of neutrino
interactions with the background particles and can be obtained using finite
 temperature and density field theory approach 
\cite{Notzold:1987ik}. The last term in above equation describes the
effective neutrino electromagnetic interactions in usual notation. 
 The flavor matrix
$\omega $ contains both electric and magnetic dipole  moments. The $V_{L}$ and
$V_{R}$ are hermitian. The $H_{T}$ evolves a system of field
equations that contain both negative and positive energy
solutions. From implications point of view, it proved convenient
to eliminate the negative energy solutions. This can be obtained
by applying the following (unitary) Foldy-Wouthuysen transformation to
$H_{T}$ and $\psi $:
\begin{equation}
U_{\rm FW}=\frac{1}{\sqrt{2}}\left(\begin{array}{cc}
                           1 & \sigma_{3} \\
                           1 & -\sigma_{3}
                           \end{array}
                     \right).
\end{equation}
    The resulting one dimensional ($x\equiv x^{3}$) Schrodinger like equations of motion 
 including the 
    effective Hamiltonian can be written as
\begin{equation}
\label{spinflip}
 \dot{\imath} \left( \begin{array}{c}
                       \dot{\nu}_{eL}\\
                       \dot{\bar{{\nu}}}_{eR} 
                  \end{array}
              \right)=\left(\begin{array}{cc}
                           0 & \mu B(r) \\
                           \mu B(r) & 0
                           \end{array}
                     \right)
                           \left( \begin{array}{c}
                       \nu_{eL}\\
                       \bar{\nu}_{eR}
                  \end{array}
                            \right),
\end{equation}
where only magnetic dipole moment, $\mu $ connecting the same
neutrino flavor in vacuum  (no neutrino interaction effects, 
 namely $V_{L}=V_{R}=V_{S}=0$) is
considered. In Eq. (\ref{spinflip}),
 the $\cdot$ denotes differentiation w.r.t distance as I use 
 $\hbar=c=1$ for relativistic neutrinos.  
Note that the strength of the magnetic
field is assumed to be varying along the neutrino trajectory.

Neutrino spin precession probability is obtained by solving
above system of equations, it is defined as $P(\nu_{eL}\to \bar{\nu}_{eR}; r)
=|\langle \bar{\nu}_{eR}(r)|\nu_{eL}(0)\rangle|^{2}$  or
\begin{equation}
\label{eight}
 P(\nu_{eL}\to \bar{\nu}_{eR}; r) = 1\cdot \sin^{2}\left(\mu
 \int^{r}_{0}B(r^{\prime}){\rm d}r^{\prime}\right),
\end{equation}
where, I have  assumed that $\nu_{eL}\, (r=0)=1$ and
  $\bar{\nu}_{eR}\, (r=0)=0$. Note that $P$ has maximal depth in
vacuum (the pre factor 1) and is independent of $E$. For $P\neq 0$,
one needs $\mu =B(r)\neq 0$, simultaneously. The neutrino spin
precession length, for a constant $B$, is defined as
\begin{equation}
 l_{B}=\pi/2\mu B,
\end{equation}
namely when the argument of $\sin^{2}$ in Eq. (\ref{eight}) is $\pi/2$,
so that if $r=l_{B}$, then $P=1$. As an elementary example, let me
ask a question in the context of Solar neutrinos:
 what $B_{\odot}$ is required to get $P=1/2$ for
$\mu \sim 10^{-11}\mu_{B}$ (assuming a constant $B_{\odot}$ in
 $0.7 \leq r/R_{\odot} \leq 1$, for simplicity) ?
Here $\mu_{B}\equiv e/2{\rm m}_{e}$ is Bohr magneton and 
 $R_{\odot}\sim 7\cdot 10^{10}$ cm is Solar radius. The
result is $B_{\odot}\sim 10$ kGauss. This is referred to as
the Voloshin, Vysotsky  and Okun (VVO) solution for the long standing
Solar neutrino problem \cite{Okun:hi}. For a review of neutrino
spin flavor oscillation solution to Solar neutrino problem, see
\cite{Akhmedov:1997yv}, whereas for a recent discussion, see
\cite{Akhmedov:2002mf}.

Let me recall here that in the standard model of particle physics with $m_{\nu}\neq 0$,
$\mu \propto m_{\nu}$ and is therefore ${\cal O}(10^{-19})\, \mu_{B}$
for $m_{\nu}\sim {\cal O}$ (1) eV. However, the $\mu $ can be as high as
$10^{-12}\, \mu_{B}$ in
some extensions of standard model \cite{Fukugita:wx}.
 It is primarily because $\mu $ dependence on $m_{\nu}$ can be changed for instance, to
 $m_{\alpha}$ (where $\alpha  =e, \mu, \tau $). In some extensions 
 of standard model, it can be achieved by 
 introducing new symmetries in the relevant standard model lagrangian density 
 (and then breaking these). The same can also be achieved either by 
 enlarging the matter particle sector or higgs or/and gauge boson sector of
 standard model.   
 The present upper bound based on measurement of $e$ 
 spectrum distortions in $\bar{\nu}_{e}e\to \bar{\nu}_{e}e$  in reactor
experiments is $\mu \, < 1.9\cdot 10^{-10}\mu_{B}$
\cite{Hagiwara:pw}.
 The interpretation of stellar cooling rate (of He burning stars) via
plasmon decay into $\nu \bar{\nu}$ imply a more stringent upper bound 
 $\mu < (1-3)\cdot
10^{-12}\, \mu_{B}$ \cite{Fukugita:1987uy}.

The neutrino spin flavor precession probability can be obtained by solving the 
following system of equations:
\begin{equation}
 \label{spinflipvacuum}
 \dot{\imath} \left( \begin{array}{c}
                       \dot{\nu}_{eL}\\
                       \dot{\bar{\nu}}_{\mu R}
                  \end{array}
              \right)=\left(\begin{array}{cc}
                           0 & \mu B(r) \\
                           \mu B(r) & \frac{\Delta m^{2}}{2E}
                           \end{array}
                     \right)
                           \left( \begin{array}{c}
                       \nu_{eL}\\
                       \bar{\nu}_{\mu R}
                  \end{array}
                            \right),
\end{equation}
which for a constant magnetic field is 
\begin{equation}
\label{vacuum}
 P(\nu_{eL}\to \bar{\nu}_{\mu R}; r) = \left[ \frac{(2\mu B)^{2}}
 {(\frac{\Delta m^{2}}{2E})^{2}+ (2\mu B)^{2}}\right]\cdot \sin^{2}\left(
 \sqrt{\left(\frac{\Delta m^{2}}{2E}\right)^{2}+ (2\mu B)^{2}}\cdot \frac{r}{2}\right).
\end{equation}
The amplitude of  $P$ is now suppressed unless $\Delta m^{2}/2E\ll 2\mu B$,
 where $\Delta m^{2}=m^{2}_{2}-m^{2}_{1}$ is the mass splitting.
This $P$ connects the neutrinos of different flavor and helicity 
 in contrast to the one given by Eq. (\ref{eight}).

A convenient form of the neutrino evolution equation that takes into account not only the
effect of neutrino interactions with matter particles in the 
 presence of external magnetic field $B$ but also mass splitting, is 
\begin{equation}
\label{adiabatic}
 \dot{\imath} \left( \begin{array}{c}
                       \dot{\nu}^{\prime}_{1}\\
                       \dot{\bar{\nu}}^{\prime}_{2}
                  \end{array}
              \right)=\left(\begin{array}{cc}
                           -(M^{2}_{2}-M^{2}_{1})/4E & -\dot{\imath}\dot{\theta}_{B}  \\
                           \dot{\imath}\dot{\theta}_{B} & (M^{2}_{2}-M^{2}_{1})/4E
                           \end{array}
                     \right)
                           \left( \begin{array}{c}
                       \nu^{\prime}_{1}\\
                       \bar{\nu}^{\prime}_{2}
                  \end{array}
                            \right).
\end{equation}
This is a form of the neutrino evolution equations that can be used for
studying numerically the propagation of mixed neutrinos \cite{Babu:1991aj}. 
 Here 
$\nu^{\prime}_{1}={\rm Exp}[-\dot{\imath}(M^{2}_{1}+M^{2}_{2})r/4E]\nu_{1}$
and 
$\nu^{\prime}_{2}={\rm Exp}[-\dot{\imath}(M^{2}_{1}+M^{2}_{2})r/4E]\nu_{2}$
 with 
\begin{equation}
 M^{2}_{2,1}=\frac{1}{2}\left[
 (m^{2}_{1}+m^{2}_{2} +V_{SF}E) \pm \sqrt{(V_{SF}E-\Delta m^{2})^{2}+(4E\mu B)^{2}}\right],
\end{equation}
where
\begin{equation}
 \tan 2\theta_{B}=\frac{2\mu B}{V_{SF}-\frac{\Delta m^{2}}{2E}}.
\end{equation}
Here, $V_{SF}\equiv \sqrt{2}G_{F}n(2Y_{e}-1)$ is the interaction potential 
 for spin flavor conversions, $n$ being the
nucleon number density. The {\tt level crossing} or resonance condition here imply
$\theta_{B}=\pi/4$ or when 
\begin{equation}
\label{level crossing}
V_{SF}=\frac{\Delta m^{2}}{2E}.
\end{equation}
Namely, in the resonance, the effective mixing angle attains its maximal value.
 In the resonance layer, the maximal change in the 
spin flavor composition of the mixed neutrino state occurs. 
The following two changes give the description for pure
neutrino flavor oscillation in appropriate basis: $\mu B \to
\frac{\Delta m^{2}}{2E}\sin 2\theta $ and $ V_{SF} \to V_{F}$ with 
 $V_{F}=\sqrt{2}G_{F}nY_{e}$.

In
Eq. (\ref{adiabatic}), if $|\dot{\theta}_{B}|\ll |(M^{2}_{2}-M^{2}_{1})/4E|$ or if
$2(2\mu B)^{2}/\pi|\dot{V}_{SF}|\gg 1$, then the off diagonal terms containing 
$\dot{\theta}_{B}$ can be ignored and $\nu^{\prime}_{1}$ and
$\nu^{\prime}_{2}$ become eigenstates of the effective
Hamiltonian. In this case, it can be shown that 
\begin{equation}
 P(\nu_{eL}\to \bar{\nu}_{\mu R}; r)=\frac{1}{2} \left(1-\cos
 2\theta^{i}_{B}\cos 2\theta^{f}_{B}\right),
\end{equation}
where $\theta^{i}_{B}$ is initial (at production) and $\theta^{f}_{B}$ is the final (at
detection) mixing angles. This situation is called the adiabatic
approximation. Note that in the adiabatic approximation, 
 the neutrino spin flavor conversion probability depends on initial and final 
 mixing angles only. If the adiabaticity is broken at the level crossing, that is, if $
\kappa_{SF}\leq 1$, where
\begin{equation}
\label{adiabaticity}
 \kappa_{SF}\equiv  \frac{2(2\mu B)^{2}}{\pi |\dot{V}_{SF}|},
\end{equation}
then one needs to solve the system of equations from the beginning.
 Note that the adiabaticity parameter $\kappa_{SF}$ does not 
 depend on $E$ explicitly. 
The violation of adiabaticity can be parameterized by 
 Landau St\"{u}ckelberg Zener probability, $P_{LSZ}$ 
as \cite{LSZ}
\begin{equation}
\label{lsz}
 P_{LSZ}= \exp{\left(-\frac{\pi^{2}}{4}\cdot \kappa_{SF}\right)}.
\end{equation}
A general expression for neutrino spin flavor 
 conversion probability including the effects of 
violation of adiabaticity in parameterized form, is
\begin{equation}
 P(\nu_{eL}\to \bar{\nu}_{\mu R}; r)=\frac{1}{2} -\left(\frac{1}{2}-P_{LSZ}\right)
 \cos2\theta^{i}_{B}\cos 2\theta^{f}_{B}.
\end{equation}
In summary, the two necessary conditions to obtain a resonant
character in neutrino oscillations are the occurrence of level crossing and fulfillment of
the adiabaticity condition at the level crossing. The 
adiabatic approximation imply $P_{LSZ} \to 0$ in the above Eq..

The details of application of above
description for neutrinos from collapsing stars is given in
\cite{Athar:1995cx}. The reader is referred to
these articles for further details. It was pointed out there that, the neutrino spin flavor
 conversions can occur for $\mu \leq 10^{-13}\, \mu_{B}$
 in a reasonable strength of magnetic field in the isotopically
neutral region of a collapsing star for 
 $10^{-1}\leq  \Delta m^{2}/{\rm eV^{2}}\leq 10^{-8}$. 
This is because of the peculiar behavior of the effective matter 
potential, $V_{SF}$ in the isotopically neutral region that a relatively small 
magnetic field strength is required to get an appreciable spin flavor 
conversion as compared to 
that in the Sun at the same distance from the center of the star.
 The above feature is possible for
spin flavor conversions between active neutrinos only \cite{Athar:pk}.

Briefly speaking, in the onion like structure of the progenitor of type II supernovae,
below the hydrogen envelope, the layers with mainly isotopically 
neutral nuclei ($n_{n}= n_{p}$) follow such as $^{4}$He, $^{12}$C, $^{16}$O, $^{28}$Si 
and $^{32}$S. Thus, the region between the hydrogen envelope and the 
core is almost isotopically neutral. The deviation from the neutrality
is the small abundance of the elements with excess of neutrons ($n_{n}\neq n_{p}$) 
 such as in $^{22}$Ne, $^{23}$Na, $^{25}$Mg and $^{56}$Fe. 
This region is referred to as isotopically neutral region, 
 it extends typically for $10^{-3}\leq r/R_{\odot}\leq 1$.

The existence of this region
follows from the fact that during the collapse, the core and inner
region of the star is neutron rich ($n_{n}>n_{p}$), whereas the outer region is
proton rich, essentially
hydrogen envelope ($n_{n}<n_{p}$),  for a typical type II supernova. 
 Obviously, the neutron and proton densities
are {\tt almost} equal in between ($n_{n}\sim n_{p}$), which defines the isotopically 
 neutral
region. In terms of $Y_{e}$, where $Y_{e}\equiv n_{e}/(n_{n}+n_{p})$, the supernova 
 phenomena imply
$Y_{e}> 0.5$ in the outer parts, whereas $Y_{e}< 0.5$ in the inner
parts (as electric neutrality of the medium implies that $n_{e}=n_{p}$). 
Therefore, $Y_{e}\sim 0.5$ in between. As $V_{SF}\propto
(2Y_{e}-1)$, this implies that $V_{SF}$ passes through very
small values in the isotopically neutral region. 
 In other words, it is {\tt suppressed} relative to $V_{F}$ 
 up to three orders of magnitude and also changes sign
 in the isotopically neutral region. 
 This is
not the case for Sun for the same distance from the center of the Sun
 because of the entirely different physics
associated with the inner parts of the Sun relative to that in a collapsing
star. The presence of the isotopically
neutral region depends on the nuclear composition of the star just after the
core collapse. It is {\tt independent} of any external $B$ present in
the expanding envelope of the collapsed star. Its presence is also
independent of the neutrino intrinsic properties such as $\Delta
m^{2}$, $\mu $ (and $\theta $).

In the case of a direct mass hierarchy ($\Delta m^{2}> 0$) and a small flavor mixing 
with $\mu =B \neq 0$, the main observational signature 
 of a neutrino spin flavor conversion is a distortion 
of the $\bar{\nu}_{e}$ flux spectrum, and specially the {\tt appearance} of a
high energy tail. In general, the final $\bar{\nu}_{e}$ spectrum 
is the energy dependent combinations of the original $\bar{\nu}_{e}$ 
spectrum and the hard spectrum of the non electron neutrinos (see Fig. 1). 
\begin{figure}
\label{figureone}
\includegraphics[width=4in]{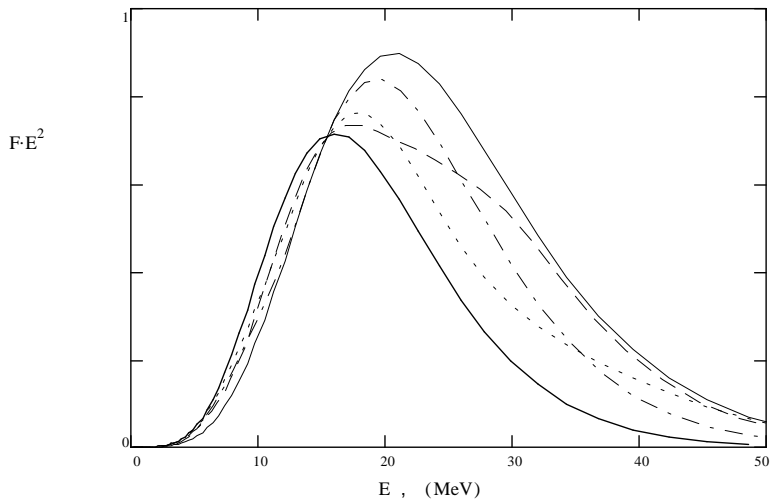}
\caption{Some examples of the $\bar{\nu}_{e}$ flux spectrum distortions in 
 neutrino spin flavor conversions. The original
 $\bar{\nu}_{e}$ flux spectrum is shown by the bold solid line. The adiabatic
 $\bar{\nu}_{e}$ conversion is shown by the solid line.}
\end{figure}
Another important signature of the spin flavor conversion can be obtained from a 
comparison of the spectra of different neutrino species. In particular,
the $\bar{\nu}_{e}$ and $\nu_{\mu}$ spectra can be completely permuted.
	
	On the other hand, in case of pure flavor adiabatic conversions, the above
mentioned features are essentially absent in case of Large Mixing Angle (LMA) 
 solution for Solar neutrino 
problem  \cite{Dighe:1999bi}. 
 Thus, the various features of future supernova neutrino data 
will possibly help to identify the role of magnetic field in the propagation 
of mixed system of neutrinos. 

	The combination of the spin-flip effects with other (flavor)
conversions may result in  rather peculiar final spectra. 
 For instance, $\nu_{e}$ may have the spectrum of the original $\bar{\nu}_{e}$, 
whereas $\bar{\nu}_{e}$ may have the original $\nu_{\mu}$ spectrum. 
The electron neutrino and anti neutrino spectra can be the same and coincide 
with the hard spectrum of the original muon neutrinos, etc.. 

Presently, the only nearby supernova from which neutrinos have been seen is SN 1987 A.
Using the fact that no $\bar{\nu}_{e}$ with average energy greater
than $\sim $ 10 MeV is seen, an upper bound on the strength of the magnetic field
profile for SN 1987A with a fixed $\mu $ value ($\mu \sim 10^{-12}\mu_{B}$) 
 was obtained under the assumption that 
$P(\nu_{\mu} \to \bar{\nu}_{e}$) does not depend on $E$ \cite{Athar:1995cx}:
\begin{equation}
 F_{\bar{\nu}_{e}}=P(\bar{\nu}_{e} \to \bar{\nu}_{e})F^{0}_{\bar{\nu}_{e}}+
 P(\nu_{\mu}\to \bar{\nu}_{e})\cdot \frac{1}{4}F^{0}_{\rm ne}.
\end{equation}
The difference among non electron neutrino spectra is ignored here. 
The upper bound is independent of any magnetic field profile inside the 
supernova as it is obtained by using Eq. (\ref{adiabaticity}) with $\kappa_{SF}=1$.
Thus, the obtained upper bound depends on the profile of $V_{SF}$ only.   
This is an example of constraining a relevant astrophysical quantity 
using neutrino observations from the core of a collapsing star.

\subsection{Prospects for future observations}
In a future nearby supernova occurring, the neutrino signal is
expected to be largely dominated by the $\bar{\nu}_{e}$ flux. This was the case
for SN 1987 A also. A reason being that 
 $\sigma (\bar{\nu}_{e}p\to n e^{+})$ is (at least) an order of magnitude
higher than the $\nu_{e}$ interaction cross section for $5 \leq
E/{\rm MeV} \leq 50$ in the detector. The $\sigma (\bar{\nu}_{e}p\to n e^{+})$ is
of the order of $10^{-41}$ cm$^{2}$ for $E\sim 25$ MeV. A future galactic 
 supernova will give several thousand neutrino events in a  Super Kamiokande
 like detector. For
completeness, in the following paragraph, I briefly summarize other
principal characteristics and limitations of present and future
detectors of neutrino bursts from supernovae.

    The light water  Cherenkov detectors such as Super Kamiokande and 
    Sudbury Neutrino Observatory (SNO)  have energy, time and angle resolution for
    $\nu_{e}$, $\bar{\nu}_{e}$ and $\nu_{\mu} \, (\nu_{\tau})$.
    The neutral current reactions in these detectors have only
    time resolution.
    The ice Cherenkov detector such as Antarctic Muon and Neutrino Detector 
 array (AMANDA) can also search for neutrinos
    from gravitational collapse.
    The heavy water Cherenkov detector (SNO) can detect
    $\bar{\nu}_{\mu}$ and $\bar{\nu}_{\tau}$ via the reaction
    $\bar{\nu}_{i}+d \to n + p+ \bar{\nu}_{i}\, \,  (E^{\rm th}=$ 2.22
    MeV) as well with time resolution. The scintillation detectors
    such as Baksan and Borexino are or will be sensitive to
    $\nu_{e}$, $\bar{\nu}_{e}$, $\nu_{\mu}$ and $\nu_{\tau}$ without
    angle resolution. The drift chamber detector, Imaging of 
 Cosmic and Rare Underground Signals (ICARUS) is or
    will be sensitive to $\nu_{\mu}(\nu_{\tau})$ with energy, time
    and angle resolution [although less sensitive to
    $\nu_{\mu}(\nu_{\tau})$]. More detailed discussion on
    prospects for future observations of supernova neutrinos can be
    found in \cite{Cei:2002mq}.

	In conclusion, all these detectors
    will in future, collectively provide the temporal, energetic,
    angular and {\tt flavor} information for any stellar collapse in our
    as well as in a nearby galaxy. This information will in turn 
     enable us to constrain the relevant astrophysical quantities 
 such as role of supernova magnetic field strength in mixed neutrino 
   propagation as elaborated in this lecture. 

\pagebreak

\section{Lecture 2: High energy neutrinos from cosmos}
\subsection{Introduction}

The neutrinos with $E > 10^{6}$ MeV are expected to mainly arise from the
interaction of ultra high energy cosmic rays considered to be protons ($p$) here 
 with the matter ($p$) and/or
radiation ($\gamma $) present in cosmos. Examples of the astrophysical sites
where these interaction can occur include the galactic plane,
other sites within our galaxy as well as distant sites such as centers of
 nearby active galaxies (AGNs) 
and cites for gamma ray bursts (GRBs).

 The plan of this lecture is to briefly review the
present motivations and status of phenomenological (and experimental) study of these 
 high
energy neutrinos. This include  a simple classification of presently 
 envisaged main sources, with a  description of the main
interactions responsible for expected high energy neutrino production. In
view of recent growing evidence of neutrino flavor oscillations,
I will elaborate the relative changes expected in the high energy neutrino
flux because of these neutrino oscillations. I will also describe
the basic crucial factors that determine the (limited) near future
prospects for observations of these high energy neutrinos. 
 Though, so far there is no observation of neutrinos with
energy greater than few thousand MeV, whose origin can not be
associated with the Atmosphere of earth, nevertheless, somewhat
optimistically speaking, given the current status of high energy
neutrino detector developments and the absolute levels of predicted high
energy neutrino fluxes, it is expected that possibly the
first evidence of high energy neutrinos may come within this decade.

A main motivation of high energy neutrino search 
is the quest of the microscopic understanding of the nature and origin
of observed ultra high energy cosmic rays, namely the presently open questions
 such as 
whether they are protons, photons, neutrinos, heavy nuclei such as
iron nuclei or some particles suggested beyond the standard
model of particle physics, and where and how they are produced or
accelerated. A positive observation of high energy neutrinos can raise
the possibility of simultaneous explanation of observed high
energy photons ($E_{\gamma}\simeq 10^{6}$ MeV) and ultra high energy 
 cosmic rays as a result of
hadron acceleration and interaction in the presently expanding
universe.

The neutrinos with energy $> 10^{6}$ MeV can act as {\tt probes}
  of the ultra high energy
phenomena observed in the Universe. Unlike photons and charged
particles such as protons and heavy nuclei, which can be absorbed or deflected by dust, 
 other
intervening matter or magnetic fields, neutrinos can more easily
reach the earth because of their weak interactions
with matter particles. It is therefore hoped that such neutrinos can
provide information about the astrophysical (or/and cosmological)  
 sources that will be
complementary to inferences based on visual observations. A better
understanding of the interactions involved in neutrino production and
a more accurate estimate of resulting neutrino fluxes could entail
important consequences. Among these are insights into 
 intrinsic properties of neutrinos  such as mass and mixing 
 \cite{Athar:2000yw}, and the possible role of gravity on
neutrino propagation in astrophysical environments \cite{Athar:2000ak}.
However, it all depends on the existence of a sizable high energy neutrino
flux.  Assuming an existence of a sizable high energy neutrino flux,
    several of the other neutrino intrinsic properties as well as the
    useful information about the source producing these neutrinos
    can be obtained, at least in principle. These include testing neutrino decay 
 hypothesis \cite{Keranen:1999nf},
    constraints on neutrino magnetic moment \cite{Mughal:hr}, 
 quantum gravity effects on
neutrino propagation \cite{Athar:1999gx}, tests of possible violation of
equivalence principle by neutrinos \cite{Minakata:1996nd}, as well as 
information on different properties of relic neutrinos
\cite{Weiler:1982qy}. Also possibly enhancement in neutrino
nucleon interaction cross section because of various new physics effects may be
constrained \cite{Tyler:2000gt}. An early attempt to constrain 
the neutrino nucleon interaction cross section is discussed in \cite{Berezinsky:kz}.

The relevant average physical picture in AGNs is as follows. 
Some galaxies have quite bright centers. The photon
luminosity of these galaxies typically reach ($10^{44}-10^{48}$)
erg s$^{-1}$. These galaxies are typically several Mpc away from us. In
 general, AGNs
refer to these bright and compact central regions, 
 which may extend up to several pc in the center. These central
compact regions have the remarkable property of being much more
luminous than the rest of the entire galaxy.  It is hypothesized
that the existence of a super massive black hole with mass, $M_{\rm BH}\,
\sim (10^{6}-10^{10})\,
 M_{\odot}$,   may explain the observed brightness as this
super massive black hole captures the matter around it through
accretion. This super massive black hole is presently hypothesized to be
formed by the collapse of a cluster of stars. Some AGNs give off a
jet of matter that stream out from the central compact
 region  in a
 transverse plane and produce hot spots when the jet strikes the
surrounding matter at its other ends. During and after accretion, the (Fermi)
accelerated protons may collide with other protons and/or with the
ambient photons in the vicinity of an AGN or/and in the
associated jets/hot spots to produce unstable hadrons. These
unstable hadrons decay mainly into neutral and charged pions. The
 neutral pions further decay dominantly into
photons and thus may explain a large fraction of the observed
brightness, 
 whereas the charged pions
mainly decay into neutrinos. AGNs, therefore,  have been targeted
as one likely source of high energy neutrinos. Currently,
the
 photohadronically ($p\gamma $) produced
flux of high energy neutrinos originating from AGNs
dominate over the flux from other sources above the relevant
Atmospheric background typically for $E\, \geq 10^{9}$ 
 MeV \cite{Protheroe:1998dm,Halzen:1998mb}.
For further reading on astrophysical super massive black holes, see
\cite{Ferrarese:2002vg}.

    Recently, fireballs are suggested as a possible production scenario 
 for gamma ray bursts as well as high energy neutrinos at the site 
\cite{Waxman:1997ti}. Though, the origin of these  gamma ray
burst fireballs is not yet understood, the observations suggest
that generically a very compact source of linear scale $\sim
10^{7}$ cm through internal or/and external shock propagation
produces these gamma ray bursts (as well as burst of high energy
 neutrinos) mainly in $p\gamma $ interactions. 
 Typically, this compact source is hypothesized
to be formed possibly due to merging of binary neutron stars or
due to collapse of a super massive star. Thus, 
fireballs have also been suggested as a probable scenario for the
observed gamma ray
 bursts, and they too are expected to emit neutrinos with energies in excess
 of hundreds of thousands of MeV. For a recent review,
see \cite{Dermer:2000yd}. 

	A nearby and more certain source of high energy neutrinos is our galactic plane.
The incoming ultra high energy cosmic ray protons interact with the ionized hydrogen 
clouds there and can produce high energy neutrinos in $pp$ interactions. Present
estimates indicate that the diffuse galactic plane muon neutrino flux can
dominate over the Atmospheric one for $E > 10^{8}$ MeV. 
\subsection{Expected neutrino production}
    A presently favorable astrophysical scenario for high energy neutrino production
 is that the
\begin{table}
\caption{Comparison of the cross sections for the three high energy neutrino production 
 interactions discussed in the text
 at $\sqrt{s}\sim 1.2\cdot 10^{3}  \, {\rm MeV}$.}
\begin{tabular}{|c|c|c|}
\hline
\hline
\label{tableone}
 Interaction & $\sigma $(mb)\\
 \hline
  $p\gamma \to N\pi^{\pm}$ & $\leq  5\cdot 10^{-1}$ \\
 $pp\to N\pi^{\pm}$ & $\sim 3\cdot 10^{1}$\\
 $\gamma \gamma \to \mu^{+} \mu^{-}$ & $ < 10^{-3}$\\
\hline
\hline
\end{tabular}
\end{table}
    observed ultra high energy cosmic rays beyond GZK cutoff (see
    later) are dominantly protons and that the observed high
    energy photon flux  can be
    associated with these. On the other hand, an unfavorable 
    scenario is that the ultra energy cosmic rays are 
 dominantly other than protons   and that
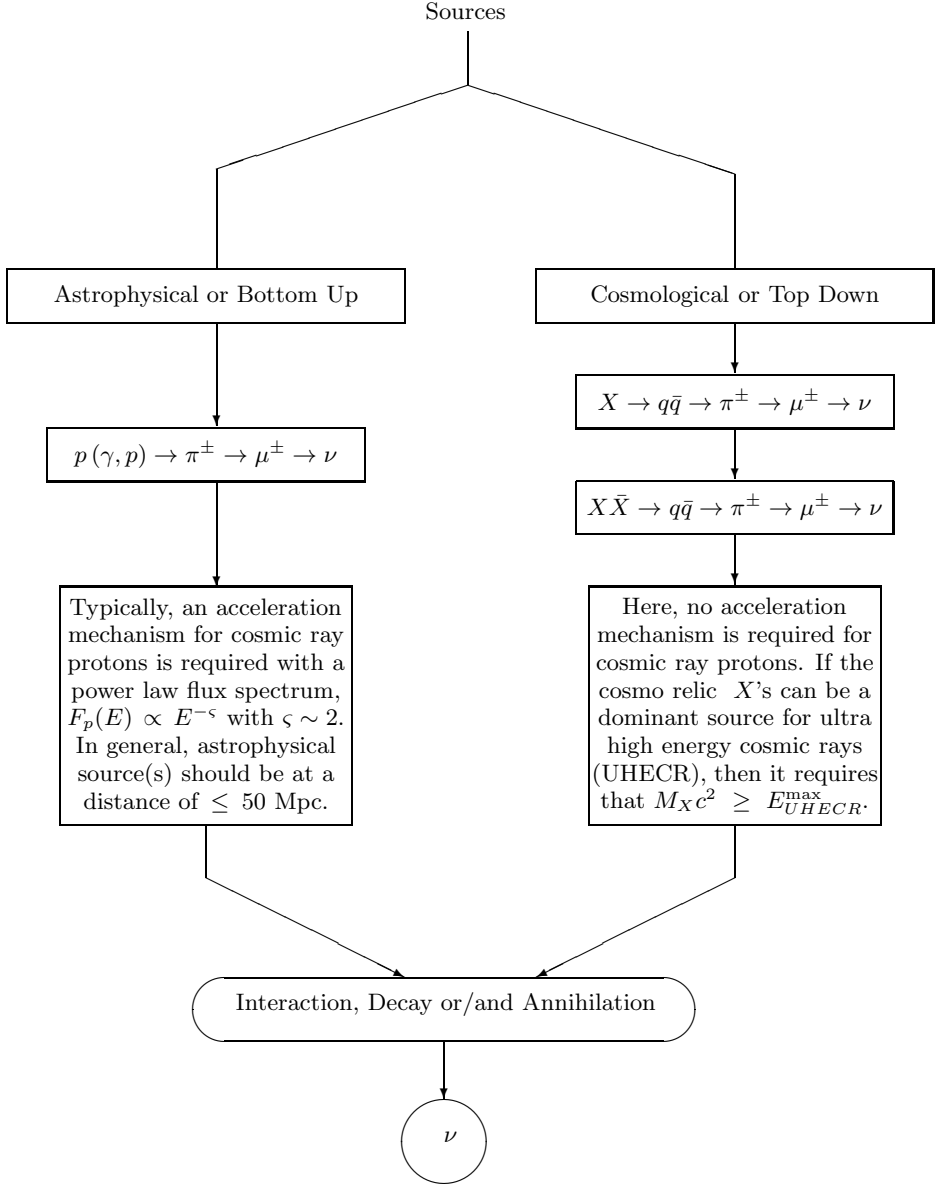
\begin{figure}
\label{figuretwo}
\begin{picture}(540,540)
\put(55,190){\framebox(110,90){\parbox{1.5in}{Typically, an acceleration mechanism 
 for cosmic ray protons is required with a power law flux spectrum, 
 $\, F_{p}(E)\, \propto \, E^{-\varsigma }$ 
 with $\varsigma  \sim 2$. 
 In general, astrophysical source(s) 
 should be at a distance of $\, \leq \, $ 50 Mpc.}}}
\put(110,190){\line(0,-1){20}}
\put(110,170){\vector(2,-1){75}}
\put(310,170){\vector(-2,-1){75}} 
\put(310,340){\vector(0,-1){19}}
\put(310,300){\vector(0,-1){19}}
\put(310,380){\vector(0,-1){19}}
\put(200,120){\oval(190,24){\makebox[1.11in][r]{Interaction, Decay or/and Annihilation}}} 
\put(200,70){\circle{30}{$\nu $}}
\put(200,108){\vector(0,-2){22}}
\put(50,320){\framebox(120,20){$p\, (\gamma,p)\to \pi^{\pm} \to \mu^{\pm}\to \nu$}}
\put(35,380){\framebox(150,20){Astrophysical or Bottom Up}}
\put(114,380){\vector(0,-1){39}}
\put(209,470){\line(-3,-1){95}} 
\put(209,470){\line(3,-1){101}}
\put(209,490){\line(0,-1){20}}
\put(310,436){\line(0,-1){36}}
\put(114,438){\line(0,-1){38}}
\put(114,320){\vector(0,-1){40}}
\put(193,495){Sources}
\put(235,380){\framebox(150,20){Cosmological or Top Down}}
\put(255,190){\framebox(110,90){\parbox{1.5in}{Here, no acceleration mechanism is
 required for cosmic ray protons.  
 If the cosmo relic $\, \, X$'s can be a dominant source for ultra high energy cosmic rays (UHECR), 
 then it requires that $M_{X}c^{2}\, \,  \geq \, \, E^{\rm max}_{UHECR}$.}}}
\put(250,340){\framebox(120,20){$X\to q\bar{q} \to \pi^{\pm}\to \mu^{\pm} \to \nu$}}
\put(310,190){\line(0,-1){20}}
\put(250,300){\dashbox{.40}(120,20){$X\bar{X}\to q\bar{q}
 \to \pi^{\pm}\to \mu^{\pm} \to \nu$}}
\end{picture}
\caption{A simple classification flow chart for presently envisaged main sources of
         high energy neutrinos. Only non tau neutrino production is illustrated.}
\end{figure}
    the observed high energy photon flux has purely
    electromagnetic origin. In the latter case, there will still
    be neutrino flux but at a rather suppressed level  
 (such as in $\gamma \gamma $ interactions) as compared
    to the former case. The latter possibility is recently
    discussed in some detail in \cite{Athar:2002ib}.

    The main interactions responsible for the production of these
    high energy neutrinos include the $p\gamma $ and $pp$  
interactions (see Table \ref{tableone}). For the behavior of these 
 cross sections as a function of center-of-mass energy 
$\sqrt{s}$ in the range of interest, see \cite{Hagiwara:pw}. There is formation of $\Delta $ 
 resonance in $p\gamma $ interactions,
 at  $\sqrt{s}\sim m_{\Delta}\sim 1.2\cdot 10^{3}$ MeV, that mainly 
decay into electron and muon neutrinos. 
 Two behaviors of the $p\gamma $ cross section, near $\sqrt{s}\sim
1.2\cdot 10^{3}$  MeV  make it an important channel for high energy neutrino
production, the relatively large width of the $\Delta $ resonance,
 $\Gamma_{\Delta}/m_{\Delta} \sim 10^{-2}$, 
and the almost constant behavior of the cross section for $\sqrt{s}> m_{\Delta}+\Gamma_{\Delta}$.
Under the assumption of all other similar 
conditions, it is the interaction cross section that determines the absolute 
level of high energy neutrino production.

 For illustrative purpose, Fig. 2 displays a simple classification 
flow chart for presently envisaged sources of high energy neutrinos. 
It includes the possibility of high energy neutrino production 
from cosmic relics, referred to as $X$ \cite{Sigl:2001th}. Briefly, these relics are 
 considered to 
be formed in the early epochs of the universe such as during inflation epoch. 
The large amount of energy trapped in these relics may be 
released in the from of grand unification scale gauge bosons
 which in turn decay/annihilate into standard model particles 
 including neutrinos. These relics need not be far away from us. In fact, 
some of the models suggest that they may be a part of our galactic dark matter halo 
 implying at a distance of $\leq $ 10 kpc. 
 If these $X$'s can be the dominant sources of
observed ultra high energy cosmic rays then this in turn severely 
constrain their number density $n_{X}$, life time $\tau_{X}$, mass $M_{X}$, and 
thus determine the resulting high energy neutrino flux spectrum shape and
absolute level. This possibility is referred to as 
the cosmological scenario for expected high energy neutrino production.
Currently, the ultra high energy cosmic rays with energy $E^{\rm max}_{UHECR}$ 
 up to $\sim 3\cdot 10^{14}$ MeV are observed \cite{Nagano:ve}.

Depending on the details of the astrophysical or cosmological
model for high energy neutrino production scenario, either the
observed photon flux or proton flux or both are used to determine the
absolute level of the expected neutrinos flux.
From the cosmos, presently high energy photons and ultra high energy cosmic
rays (considered to be protons here) are observed in the relevant context. 
 Their observed level of flux  determines the absolute 
 flux level of neutrinos as high energy neutrinos are secondary in nature in the
sense that they are not matter particles and are not a significant
 fraction of the matter density associated with a specific known 
 astrophysical or/and cosmological  source. On
the other hand, neutrinos are stable and neutral and therefore for
this precise reason will carry useful information about the
source. Supposing protons can escape the {\tt extra galactic} astrophysical sources and  
can be a dominant fraction of the observed ultra high energy cosmic ray
flux, the resulting high energy (muon) neutrino flux mainly in $p\gamma $ and $pp$ interactions 
either arising from inside the source or during propagation has to be less than
 this. It can be typically $\leq 10^{-5}\, {\rm MeV}({\rm cm}^{2}\cdot {\rm s \cdot sr})^{-1}$ 
  for
 $10^{8}< E/{\rm MeV} < 10^{15}$ \cite{Waxman:1998yy}. This bound further
 tightens by a factor of 1/2 once the neutrino flavor oscillation effects  are taken 
into account (see later).

	Consider now briefly the $p\gamma \, \to  \Delta \to p\pi\, (N=p)$ interactions 
	 occurring during 
 the propagation of ultra high energy cosmic rays either inside an 
 astrophysical source or between the source and the earth 
in the presence of  a dense photon background. This is to serve as an illustrative example
 for having an order of magnitude idea of the expected $E$. 
 The threshold energy for protons interacting at an angle $\phi $ to form $\Delta $ resonance, is
\begin{equation}
    E^{\rm th}_{p}=
    \frac{(m_{p}+m_{\pi})^{2}-m_{p}^{2}}{2E_{\gamma}(1-\cos\phi)},
\end{equation}
which in case of head on interactions further simplifies to
\begin{equation}
\label{GZK}
    E^{\rm th}_{p}\simeq \frac{m_{p}m_{\pi}}{2E_{\gamma}}.
\end{equation}
 For $E_{p}<E^{\rm th}_{p}$, the interaction $p\gamma \to p e^{+}e^{-}$
 dominates the energy loss for protons. If $E_{\gamma}=E^{\rm CMB}_{\gamma}\sim 2.7$ K then 
$E^{\rm th}_{p}\sim  10^{14}$ MeV. 
 The $p\gamma $ interaction length can be defined
as
\begin{equation}
\lambda \sim 1/n_{\gamma}\sigma_{p\gamma \to p\pi}.
\end{equation}
For instance, if 
 $n_{\gamma}=n^{\rm CMB}_{\gamma} \sim 410$ cm$^{-3}$ for $E_{\gamma}=E^{\rm CMB}_{\gamma}$ 
 then $\lambda < 6$ Mpc, where $\sigma_{p\gamma \to p\pi}$ is given in Table I.

The propagation of ultra high energy proton flux, $F_{p}$ can be studied in 
 the presence of photon background in distance $r$, by
solving the following equation
\begin{equation}
    \frac{{\rm d}F_{p}}{{\rm d}r}=-\frac{1}{\lambda}F_{p}.
\end{equation}
The negative sign indicates the decrease in the ultra high energy proton
flux because of interaction described by $\lambda $. 
 This results in an exponential cut off in $p$ flux spectrum. In case of 
 ubiquitous cmb photon background, it is
 commonly referred to as Greisen Zatsepin Kuzmin (GZK) cut off \cite{Greisen:1966jv}. 
It occurs at $E^{\rm th}_{p}\sim 10^{14}$ MeV, according to Eq. (\ref{GZK}). 
The 
resulting GZK (muon) neutrino flux spectrum peaks at $\sim 10^{12}$ MeV by sharing 
roughly (1/4)$\cdot $(1/5) of the $E^{\rm th}_{p}$. 

The matter density in interstellar medium as well as in several of the astrophysical 
sites such as the galactic plane, the AGNs and the GRBs, is rather small 
 (relative to that in Atmosphere of 
earth). Therefore, a rather simple formula can be used to estimate high 
energy neutrino flux spectrum in $p\gamma $ and/or $pp$ in a 
specific individual astrophysical site
\begin{equation}
\label{astroflux}
 F^{0}_{\nu}(E)=\int^{E^{\rm max}}_{E}
  {\mbox d}E\,  F_{p}\, (E)\,  g(E)
 \frac{\mbox{d}n_{p(\gamma ,p) \to \nu Y}}   {\mbox{d}E}.
\end{equation}
Here $F_{p}(E)$ parameterizes the high energy proton
flux. The function $g(E)\equiv r/\lambda
$ gives the number of $p (\gamma, p)$ interactions within the
 distance $r$. The ${\rm d}n/{\rm d}E\equiv \sigma^{-1}{\rm
d}\sigma/{\rm d}E$ is the neutrino energy distribution in above interactions.
The implicit assumption here is that the unstable hadrons and leptons produced 
in above interactions decay before they interact owing to the fact that the matter
density in the distance $r$ is assumed to be rather small. Also, the effects of possible 
 red shift
evolution and magnetic field  of the astrophysical sources are neglected for simplicity.

There is yet another possible class of astrophysical sources of high energy neutrinos
that are essentially neither constrained by observed high energy photon nor by ultra high
energy cosmic ray flux. It is so because in this class of sources, the
matter density is considered to be too large so that neither of the above
leave the source. These sources are therefore commonly referred to as hidden sources or 
 neutrinos only sources.
 The high energy neutrino production occurs in same $pp$ (or $p\gamma $) interactions here 
  also.  
 These can only be constrained by the high energy neutrino flux
(non) observations \cite{Berezinsky:2000bq}.

The above discussion is restricted to non tau neutrino production only. 
 In the $\pi^{\pm}\to \mu^{\pm} \to \nu$ 
decay situation, 
 the relative ratio of resulting electron and muon neutrino flux is 1 : 2 respectively. 
 The astrophysical tau neutrino flux is produced in decays of $D^{\pm}_{S}$. For 
 $\sqrt{s}\sim m_{\Delta} $, it is known that 
 $ \sigma[p(\gamma ,p)\to D^{\pm}_{S}Y]/ \sigma[p(\gamma ,p)\to \pi^{\pm}Y] 
 \leq {\cal O}(10^{-3}-10^{-4})$. 
 The high energy tau neutrino flux is thus rather suppressed at the production 
sites and can therefore be taken as approximately zero, 
 resulting in 1 : 2 : 0 \cite{Athar:1998ux}.
 For a recent review on astrophysical tau neutrinos, see \cite{Athar:2002rr},
whereas for cosmological tau neutrinos, see, for instance \cite{Wichoski:1998kh}.

\subsection{Oscillations during propagation: Effects of neutrino mixing}

There are at least two aspects of neutrino propagation effects
that need somewhat careful considerations in study of neutrino 
 mixing effects for high energy neutrinos. These are: the neutrino 
interactions with the background particles inside the (astrophysical) source of
neutrinos as well as between the source and the earth.  
The present knowledge of matter density, $\rho $
inside the known sources as well as between these sources and the
 earth imply that it is rather quite small (as compared to that in Sun). 
 As a  result, the level crossing
condition, $G_{F}\rho/m_{N}\sim \Delta m^{2}/2E$, for  matter enhanced neutrino flavor 
 oscillations is 
 not satisfied  [see also, for instance, Eq. (\ref{level crossing})]. 
 Level crossing is a necessary condition for occurrence of matter 
 enhanced neutrino flavor oscillations.  Therefore, there are
essentially no matter effects on pure vacuum flavor oscillations.
Note that this is in contrast to the situation in supernovae.  
 Furthermore, the neutrino nucleon and neutrino electron inelastic 
 interaction effects are also small enough to effect the mixed neutrino propagation even
at ultra high energy in a significantly observable manner. 
 This is also because of rather small matter density.   
 Therefore, I elaborate only effects of neutrino flavor 
mixing in vacuum (with no matter interactions)\footnote{If 
i) $0.1\leq \sin^{2}2\theta \leq 0.95 $, ii) $E\geq 10^{12}$ MeV, 
iii) the red shift $z\geq 3$ at production, and iv) $\xi \geq 1$, where 
 $\xi\equiv (n_{\nu}-n_{\bar{\nu}})/n_{\gamma}$, then a deviation from pure 
vacuum flavor oscillations can be of the order of few percent, when high energy 
 neutrinos
scatter over the very low energy relic neutrinos during their propagation to us 
 in the interstellar medium \cite{Lunardini:2000fy}.}.

Note from the previous subsection that the high 
 energy neutrinos
are produced in the following relative ratios
\begin{equation}
\label{initial}
F^{0}_{\nu_{e}}: F^{0}_{\nu_{\mu}}: F^{0}_{\nu_{\tau}}=1: 2: 0.
\end{equation}
It is assumed here that the high energy neutrinos and anti neutrinos originate
in equal proportion from a source and are counted in the symbol $\nu $ together. 
 Also as the absolute level of 
high energy neutrino flux is presently unknown, I therefore elaborate
the neutrino mixing effects on relative ratios only, in the
context of three flavors. 
 Four flavor mixing effects are considered in 
 \cite{Athar:2000yw,Athar:2000tg}.

To obtain a general expression for flavor oscillation formula, I
start with the connection $U$ between the flavor $|\nu_{\alpha}\rangle $
 and mass $|\nu_{i}\rangle$ eigen 
states of neutrinos, namely
\begin{equation}
|\nu_{\alpha}\rangle =\sum^{3}_{i = 1}U_{\alpha i}|\nu_{i}\rangle,
\end{equation}
where $\alpha = e, \mu $ or $\tau $. 
 In the context of three neutrinos, $U$ is
called Maki Nakagawa Sakita (MNS) mixing matrix \cite{Maki:mu}.
  It can be
obtained by performing the following operations to coincide with the 
one given in \cite{Hagiwara:pw}: 
\begin{equation}
 U\equiv R_{23}(\theta_{23})\cdot {\rm diag}(e^{-i\delta_{13}/2},
 1,\, e^{i\delta_{13}/2})\cdot R_{13}(\theta_{13})\cdot
 {\rm diag}(e^{i\delta_{13}/2},1,\, e^{-i\delta_{13}/2})
 \cdot R_{12}(\theta_{12}),
\end{equation}
where $\theta$'s are neutrino mixing angles and $\delta_{13}$ is CP violation phase.
Explicitly, it reads
\begin{equation}
 U=\left( \begin{array}{ccc}
          c_{12}c_{13} & s_{12}c_{13} & s_{13}e^{-i\delta_{13}}\\
          -s_{12}c_{23}-c_{12}s_{23}s_{13}e^{i \delta_{13}} & 
          c_{12}c_{23}-s_{12}s_{23}s_{13}e^{i\delta_{13}} & 
          s_{23}c_{13}\\
          s_{12}s_{23}-c_{12}c_{23}s_{13}e^{i\delta_{13}} & 
          -c_{12}s_{23}-s_{12}c_{23}s_{13}e^{i\delta_{13}} & 
          c_{23}c_{13}
          \end{array}
   \right).
\end{equation}
Here $c_{ij}=\cos \theta_{ij}$ and $s_{ij}=\sin \theta_{ij}$ (with $j=1,2,3$) 
 and that $UU^{\dagger}=1$.
Using it, one obtains the following well known formula for flavor
 oscillation probability from $\alpha$ to $\beta \, (\beta =e, \mu \, {\rm or}\, \tau) $ 
 for a neutrino source at a fixed distance $L$: 
\begin{equation}
\label{complete}
P(\nu_{\alpha} \to \nu_{\beta}; L) \equiv P_{\alpha \beta}=  
 \delta_{\alpha \beta}-\sum_{j\neq k}
 U^{*}_{\alpha j}U_{\beta j}U_{\alpha k}
 U^{*}_{\beta k} (1-e^{-i\Delta m^{2}_{jk}L/2E}). 
\end{equation}
In the far distance approximation, namely, in the limit $L\to \infty$, one obtains
\begin{eqnarray}
\label{osc-prob}
 P(\nu_{\alpha} \to \nu_{\beta}; L\to \infty)& \simeq & 
 \delta_{\alpha \beta} -\sum_{j \neq k} 
 U^{*}_{\alpha j}U_{\beta j}U_{\alpha k}
 U^{*}_{\beta k},\nonumber \\ 
  & \simeq  & 
 \sum^{3}_{j =1} |U_{\alpha j}|^{2}
 |U_{\alpha j}|^{2}.
\end{eqnarray}
Because of the assumed averaging over the rapidly oscillating phase
 ($l_{\rm osc}\ll L, {\rm where} \, \, l_{\rm osc}\equiv 2E/\Delta m^{2}_{jk}$), 
  the last two expressions
 are {\tt independent} of $E$  and $\Delta m^{2}$. Under this assumption, the oscillation 
probability can be written as a symmetric matrix $P$ and $P$ can be written as
a product of a matrix $A$:
\begin{equation}
 P=\left( \begin{array}{ccc}
          P_{ee} & P_{e\mu } & P_{e\tau}\\
          P_{e\mu } & P_{\mu \mu } & P_{\mu \tau}\\
	    P_{e\tau } & P_{\mu \tau} & P_{\tau \tau} 
          \end{array}
   \right)\equiv AA^{T},
\end{equation}
with
\begin{equation}
  A=\left( \begin{array}{ccc}
          |U_{e1}|^{2} & |U_{e2}|^{2} & |U_{e3}|^{2} \\
          |U_{\mu 1}|^{2}  & |U_{\mu 2}|^{2}  & |U_{\mu 3}|^{2} \\
	    |U_{\tau 1}|^{2}  & |U_{\tau 2}|^{2}  & |U_{\tau 3}|^{2} 
          \end{array}
   \right).
\end{equation}
A  simple form for $P$ matrix can be obtained  in case of
vanishing $\delta_{13} $ and $\theta_{13}$  with bi maximal mixing \cite{Athar:1999at}:
\begin{equation}
 P=\left( \begin{array}{ccc}
          5/8  & 3/16  & 3/16\\
          3/16 & 13/32 & 13/32\\
          3/16 & 13/32 & 13/32
          \end{array}
   \right).
\end{equation}
This $P$ matrix satisfies the following unitarity
conditions:
\begin{equation}
 1-P_{ee}=P_{e\mu}+P_{e\tau}, \, \, \, 
 1-P_{\mu \mu}=P_{e\mu}+P_{\mu \tau}, \, \, \, 
 1-P_{\tau \tau}=P_{e\tau}+P_{\mu \tau},
\end{equation}
namely, the disappearance of a certain neutrino flavor is equal to 
the appearance of this flavor into other (active) neutrino flavors. 
High energy neutrino flux arriving at the earth can be estimated 
using 
\begin{equation}
\label{atearth}
 F_{\nu_{\alpha}}=\sum_{\beta}P_{\alpha \beta}F^{0}_{\nu_{\beta}},
\end{equation}
where $P_{\alpha \beta}$ is given by Eq. (\ref{osc-prob}). Note that in case
of initial relative flux ratios as 1 : 2 : 0 [see Eq. (\ref{initial})], one always get 
\begin{equation}
 F_{\nu_{e}}: F_{\nu_{\mu}}: F_{\nu_{\tau}}=1: 1: 1,
\end{equation}
under the assumption of
averaging irrespective of any specific flavor oscillation 
 solution for Solar neutrino problem \cite{Athar:2000je}.
 A considerable enhancement in $F_{\nu_{\tau}}$ 
 relative to $F^{0}_{\nu_{\tau}}$ because of neutrino oscillations is evident.   
  A some what detailed numerical study that takes into account 
 the effects of non vanishing $\delta_{13}$ and $\theta_{13}$ indicates that the 
 deviation from these 
final relative ratios is not more than few percent 
 (namely, $|\epsilon | \leq 0.1$ in $1\pm |\epsilon |$) \cite{Athar:2000yw}.   
  There could, in principle, be several intrinsic neutrino properties that may
lead to {\tt deviations} from 1 : 1 : 1 final relative ratios other than $|\epsilon |$ as well as 
 an {\tt energy dependence}, such as neutrino spin flavor conversions \cite{Athar:1999gw}. 
 Astrophysical/cosmological reasons at the source can also contribute to these deviations.

In the  above simplified discussion, the expression for $P$ neither depends on $\Delta
m^{2}$ nor on $E $. However, in some situations, this need not be the case. 
 In that case, one need to use complete expression for
$P$ given by Eq. (\ref{complete}) and have to average over the red shift 
 distribution of astrophysical sources, $f(z)$. 
 This gives the effect of
evolution of the sources with respect to $z$. This effect can be calculated
using the $P$ given in Eq. (\ref{complete}) with $E \to (1+z)E$ in 
following formula
\begin{equation}
 P_{\alpha \beta}(E)=\frac{\int^{z_{max}}_{0}P_{\alpha \beta}(E,z) f(z) {\rm d}z}
 {\int^{z_{max}}_{0} f(z) {\rm d}z}.
\end{equation}
The $f(z)$ can be found in \cite{Lunardini:2000fy}.	
\subsection{Prospects for possible future observations}
The current status of the dedicated high energy neutrino detectors
is given in \cite{resvanis}. Briefly, 
 the detectors based on Cherenkov radiation measurement, in ice or water are 
 the Antarctic Muon and Neutrino Detector Array (AMANDA) and 
 its proposed extension, the Ice Cube, the lake Baikal detector and the Astronomy with 
 a Neutrino Telescope and Abyss environmental RESearch (ANTARES) detector array. 
The hybrid detectors based on particle and radiation measurement such as Pierre 
Auger Observatory can also detect high energy neutrinos \cite{Capelle:1998zz}. 
 Detectors based on alternative
detection techniques such as radio wave detection are also in operation, such as
Radio Ice Cherenkov Experiment (RICE). This detector is based 
 on Askaryan effect. This effect is briefly defined as follows: 
 In an electromagentic shower generated in deep inelastic neutrino nucleon 
 interaction, the electrons and photons in the shower generate an excess 
 of $\sim 10-20 \, \%$ electrons in the shower because of the electron and 
 photon interactions with the medium in which the shower develops. 
 This in turn generate  coherent radio wave pulse (in addition to other type 
 of radiation), if the wavelength of this radio emission is greater 
 than the size of the shower.
	The search for alternative high energy neutrino detection 
 medium other than air, water and/or ice, such as rock salt  
 has also been attempted for radio wave emission \cite{Chiba:2000fj}. 
 It might also be possible to detect the
 acoustic pulses generated by deep inelastic neutrino nucleon interactions near or inside the 
 detector. An attempt in this direction is through 
 Sea Acoustic Detection of Cosmic Objects (SADCO) detector array.
  Other proposals include space based detectors such as Orbiting Wide Angle Light collector 
 (OWL/Air Watch)
 and Extreme Universe Space Observatory (EUSO).

 Among all these, the tightest upper bounds on high energy neutrino flux are reported by 
 AMANDA and Baikal 
detectors. From AMANDA, this is typically 
 $\leq 10^{-3}\, {\rm MeV} ({\rm cm}^{2}\cdot {\rm s}\cdot {\rm sr})^{-1}$ in the energy
range $10^{6}< E/{\rm MeV} < 10^{9}$.  The effective area for AMANDA detector 
 is $\sim 10^{-2}$ 
 km$^{2}$ for a $10^{7}$ MeV muon neutrino. The next generation high energy neutrino 
 detectors are considered to have an effective area of $\sim $ 1 km$^{2}$.

 The high energy neutrino observation can be achieved in the
 following two main interactions: the deep inelastic neutrino nucleon and neutrino
 electron interactions. The deep inelastic neutrino nucleon interaction can 
 proceed via $Z$ or $W^{\pm}$ exchange. The former is 
 called neutral current (NC) interaction, whereas the later is called charged current (CC)
 interaction. The CC interactions $(\nu_{\alpha}N\to \alpha Y)$ are most relevant for 
  prospective high 
 energy neutrino observations. The showers, the charged particles and the 
 associated radiation emission such as Cherenkov radiation from these interactions 
 are the measurable  quantities. 
 The CC deep inelastic neutrino
 nucleon cross section $\sigma^{CC}(E)$ over nucleons  with mass $m_{N}$,  
 can be written as a function of  the incoming neutrino energy $E$ as:
\begin{eqnarray}
\label{DIS}
 \sigma^{\rm CC}(E) & = & \frac{2G^{2}_{F}m_{N}E}{\pi}\int {\rm d}x \int {\rm d}y
 \left (\frac{M^{2}_{W}}{Q^{2}+M^{2}_{W}}\right)^{2} \cdot x \cdot
 \nonumber \\
 & &
 \left \{ \left(1-\frac{m^{2}_{N}x^{2}y^{2}}{Q^{2}}\right)
 [f_{d}(x, Q^{2})+f_{s}(x, Q^{2})+f_{b}(x, Q^{2})]+ \right. \nonumber \\
 & & \left. \left( (1-y)^{2}-\frac{m^{2}_{N}x^{2}y^{2}}{Q^{2}}\right)
 [f_{\bar{u}}(x, Q^{2})+f_{\bar{c}}(x, Q^{2})+f_{\bar{t}}(x, Q^{2})]\right\}.
\end{eqnarray}
The integration limit for $x$ and $y$ can be taken between 0 and 1. 
This expression can be straight forwardly obtained using $s=2m_{N}E$ 
 in \cite{Hagiwara:pw}. Here $f_{i}(x, Q^{2})$
are the parton distribution functions. In above Eq., $y\equiv (E-E^{\prime})/E$ is the 
 inelasticity in the neutrino nucleon interactions. 
 It gives the fraction of $E$ 
lost in a single neutrino nucleon interaction. 
 The $x \equiv Q^{2}/2m_{N}(E-E^{\prime})$ is the fraction of 
 the nucleon's momentum carried by the struck quark.
 The lepton mass is ignored here in comparison with $m_{N}$.

\begin{figure}
\includegraphics[width=2in]{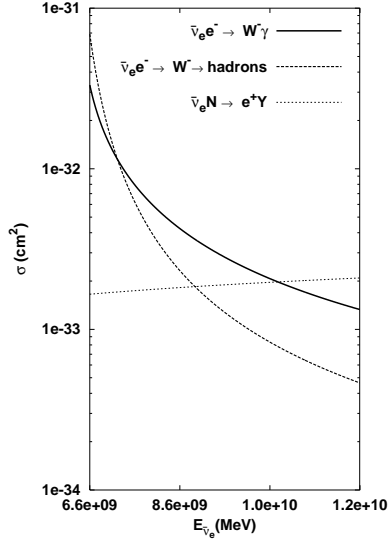}
\caption{Examples of high energy $\bar{\nu}_{e}$ absorption cross section 
 over two different target particles as a function of $E_{\bar{\nu}_{e}}$ (MeV).
 The minimum value of $E_{\bar{\nu}_{e}}$ corresponds to $(M_{W}+\Gamma_{W})^{2}/2m_{e}$.}
\label{fig}
\end{figure}

 The neutrino electron interaction cross section 
 on the other hand has a resonant character for $s=2 m_{e}E_{\bar{\nu}_{e}}$:
\begin{equation}
\label{sigma}
 \sigma(\bar{\nu}_{e}e \to W^{-} \to {\rm hadrons})
 =\frac{\Gamma_{W}({\rm hadrons})}{\Gamma_{W}({e^{+}\nu})}\cdot \frac{G^{2}_{F}s}{3\pi}
 \cdot \left[\frac{M^{4}_{W}}
 {(s-M_{W}^{2})^{2}+\Gamma^{2}_{W} M^{2}_{W}}\right],
\end{equation}
where $\Gamma_{W}$'s can be found in \cite{Hagiwara:pw}. The
above resonant interaction select the anti electron neutrino flavor as well as the
energy. This interaction can, in principle, be used to calibrate the 
 high energy neutrino energy  provided it can possibly be discriminated from neutrino nucleon 
 interaction in a detector. The 
 $\sigma(\bar{\nu}_{e}e \to W^{-} \to {\rm hadrons})$
 has a slight enhancement because of hard photon emission in the 
final state for $\sqrt{s}\geq \Gamma_{W}$. It is given by \cite{Athar:2001bi}
\begin{equation}
\label{six}
 \sigma(\bar{\nu}_{e}e^{-}\rightarrow W^{-}\gamma)=
 \frac{\sqrt{2}\alpha G_{F}}{3u^{2}(u-1)} \left[3(u^{2}+1)\ln
 \left\{\frac{(u-1)M^{2}_{W}}{m^{2}_{e}}\right\}
       -(5u^{2}-4u+5)\right],
\end{equation}
where $u=s/M^{2}_{W}$. 
In Fig. \ref{fig}, the three cross sections are plotted for illustration. 
The $\sigma^{\rm CC}(\bar{\nu}_{e}N)$ is calculated using CTEQ(5M) parton 
 distribution functions generated by Coordinated Theoretical and Experimental Project 
 on QCD Phenomenology and Tests of the Standard Model \cite{Lai:1994bb}.

The high energy neutrino flux arrives at an earth based detector in three 
general directions in equal proportion. 
The downward going neutrinos do not cross any significant earth cord before reaching
the detector. The horizontal and upward going neutrinos cross the earth with increasing 
 cord length respectively.

The event rate for downward going high energy neutrinos in CC deep 
inelastic interactions is given by \cite{Gandhi:1998ri}
\begin{equation}
    {\rm Rate}= A \int^{E^{\rm max}_{\nu_{\alpha}}}_{E^{\rm min}_{\alpha}} {\rm
    d}E_{\nu_{\alpha}} P_{\nu_{\alpha}\to \alpha} (E_{\nu_{\alpha}}, E^{\rm min}_{\alpha})
    F_{\nu_{\alpha}},
\end{equation}
here $A$ is the area of the high energy neutrino detector. 
 The $F_{\nu_{\alpha}}$ can be obtained using Eq. (\ref{atearth}). In the above
equation
\begin{equation}
\label{prob}
 P_{\nu_{\alpha }\to \alpha } (E_{\nu_{\alpha }}, E^{\rm min}_{\alpha })= N_{A}
  \int^{1-E^{\rm min}_{\alpha }/E_{\nu_{\alpha }}}_{0}
  {\rm d}y R_{\alpha }(E_{\nu_{\alpha }}, E^{\rm min}_{\alpha })
  \frac{{\rm d}\sigma^{\rm CC}_{\nu_{\alpha }N}(E_{\nu_{\alpha }}, y)}{{\rm
  d}y}.
\end{equation}
The ${\rm d}\sigma/{\rm d}y$ can be obtained using Eq. (\ref{DIS}).  
 The $N_{A}$ is the Avogadro's constant. Various $R$'s are given in
Table \ref{table2}. Note that for $\tau $ lepton,
 it is the decay length that is considered as its range
 with $E^{\rm min}_{\tau}\sim 2\cdot 10^{9}$ MeV and 
 $E^{\rm max}_{\nu_{\tau}}\sim 2\cdot 10^{10}$ MeV as
 the value of $D$ is chosen as 10$^{5}$ cm for illustration 
 here \cite{Athar:2000rx}.
Also note that $R_{e}\equiv R_{e}(E)$ only.

 For upward going high energy neutrinos, a shadow factor $S$
is included in the integral given by Eq. (\ref{prob}). 
\begin{table}[b]
\caption{The three charged lepton ranges discussed in the text.}
%\label{table2}}
\label{table2}
\begin{tabular}{|c|c|c|}
\hline
\hline
Lepton Flavor & $R $(cmwe) & Ref. \\
\hline
 $e$& 40 $\left[(1-\langle y(E) \rangle) \frac{E}{6.4\cdot 10^{7}{\rm MeV}}\right]$ 
 & \cite{Gandhi:1998ri}\\
 $\mu$ & $\frac{1}{b}\ln \left( \frac{a+bE_{\mu}}{a+bE^{\rm min}_{\mu}}\right)$,
 $a = 2\cdot 10^{-3}\, \, \,  10^{3}$MeV/cmwe, b =  $3.9\cdot 10^{-6}$ /cmwe & 
 \cite{Gandhi:1998ri}\\
 $\tau$ & $D - \frac{E(1-y)\tau_{c}}{m_{\tau}c^{2}}$, D = $10^{5}{\rm cm}$ & 
 \cite{Athar:2000rx}\\
\hline
\hline
\end{tabular}
\end{table}
The shadow factor $S$ takes into account the effects 
 of absorption by earth \cite{Gandhi:1998ri}.
 The absorption of upward going high energy neutrinos by earth is
 {\tt neutrino flavor
dependent}. For $E \, \geq 10^{9}$ MeV, the upward  going tau
neutrinos
 may reach the
surface of the earth in a relatively small number by lowering their energy so that
 $E \, < 10^{9}$ MeV \cite{Halzen:1998be},
whereas the upward going electron and muon neutrinos are almost
completely
 absorbed by the earth. 
 For further details,
see \cite{Gandhi:1998ri}.  
 For downward going high energy neutrinos, the $S$ is taken as unity.  
\begin{figure}[t]
\hglue -8.cm
\includegraphics[width=4in]{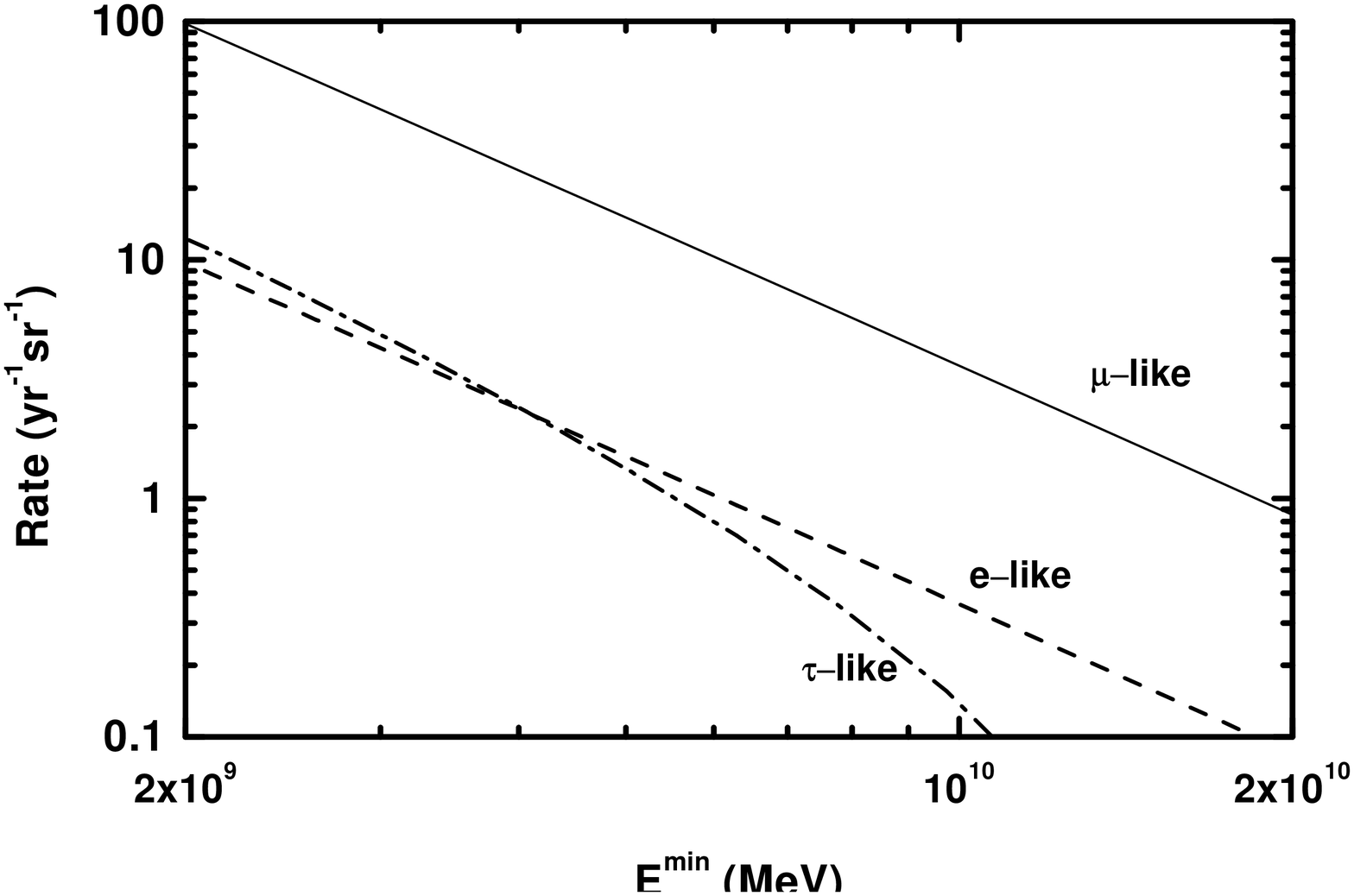}
\vglue -7.0cm 
\hglue 8.8cm \includegraphics[width=2.8in]{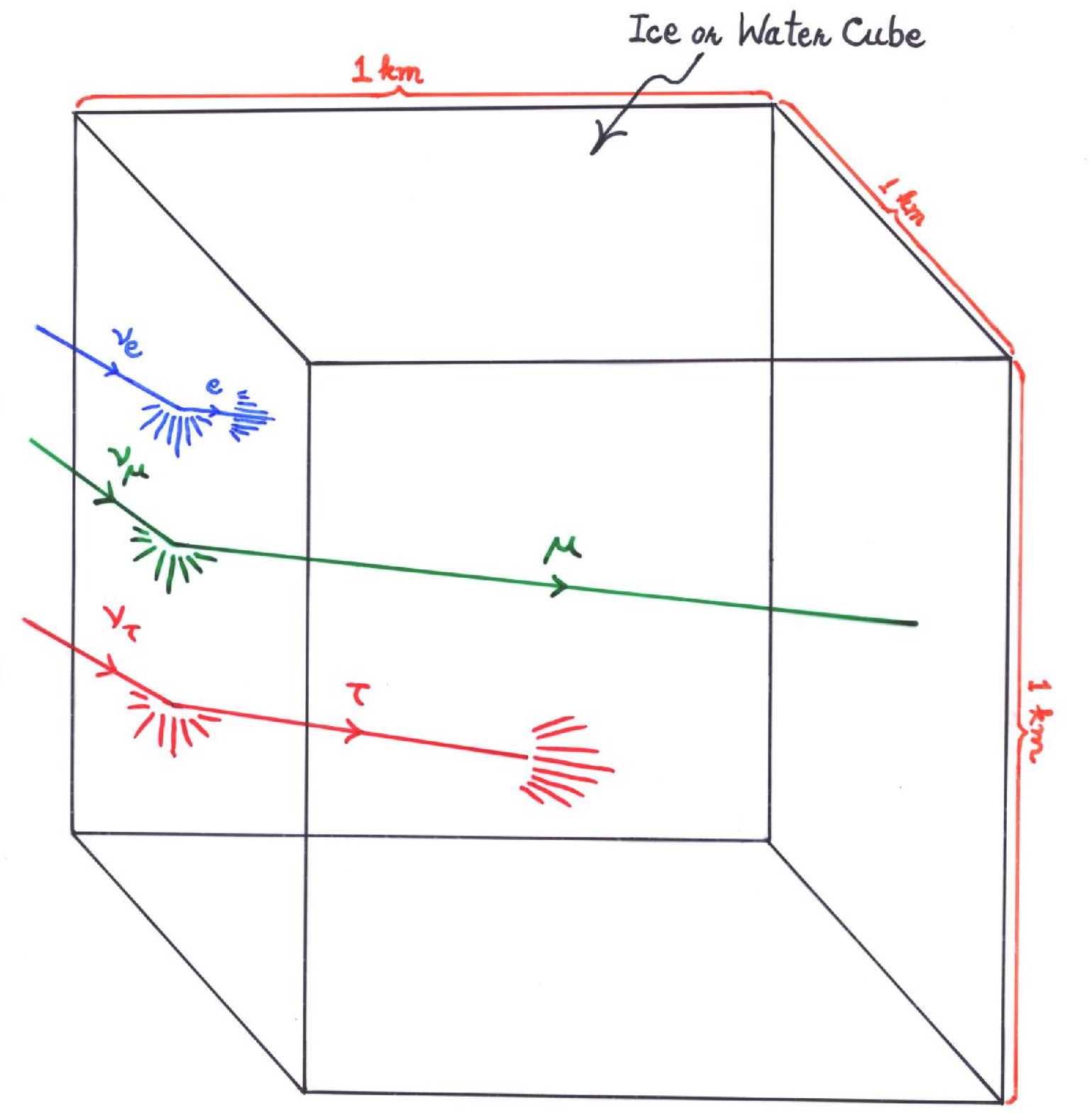}
\caption{Left panel: Expected downward going $e-$like,
 $\mu-$like and $\tau-$like event rate produced by AGN neutrinos as
 a function of minimum energy of the corresponding charged lepton
 in a large km$^{3}$ volume ice or water neutrino detector. Three
 flavor neutrino mixing is assumed.
 Right panel: 
 Approximate representative event topologies for the three neutrino 
 flavors in a 
 km$^{3}$ volume water or ice neutrino detector for the order of 
 magnitude energy interval shown in left panel.}
\label{eventrate}
\end{figure}
Detailed estimates of the high energy neutrino  event rates are
done mainly numerically. 
 These estimates  are model
dependent. The event rates of  downward high energy 
neutrinos  typically vary between $\sim \, {\cal O}(10^{1})$
and
 $\sim \, {\cal O}(10^{2})$ in units of (yr  sr)$^{-1}$ for the proposed 
 km$^{3}$ volume ice or water neutrino detector.
The left panel of 
 Fig. \ref{eventrate} displays the three downward going 
event rates along with examples of event topologies for  AGN neutrinos  \cite{Szabo:qx}.
 In this AGN model, the $pp$ interactions inside the core of the AGN are  
 considered to play an important role. 
The $e-$ like event rate are obtained by re scaling the
$\mu-$ like event rate. The indicated order of magnitude 
energy interval is relevant for proposed km$^{3}$ volume high energy 
 neutrino detectors in the context of possible neutrino flavor 
 identification.

The downward going  high energy neutrinos of different
flavors interact with the medium (free nucleons) of the detector,
deep inelastically mainly through CC 
interactions. The three flavors on the average give rise to {\tt
different} event topologies
 based on
these interactions and the behavior of the associated 
{\tt charged lepton}.
 For instance, for $E \, \geq 10^{9}$ MeV, in proposed km$^{3}$ volume
 ice or water  
 neutrino detectors, typically the downward going high energy
 electron neutrinos produce a single shower, the muon neutrinos
produce muon like tracks passing through the detector (along with a single shower), 
 whereas the
 tau neutrinos produce two
 hadronic showers connected by muon like track and is such that the amplitude
  of the second shower is essentially a factor of two larger as compared to the first.
 Here, amplitude refers to maximum number of charged particles per unit
 length (see right panel of Fig. \ref{eventrate}) \cite{Learned:1994wg,Athar:2000rx}. 
 For a recent discussion on prospects for observations of 
near horizontal (tau)  neutrinos, see \cite{Feng:2001ue}.

\section{Summary and Conclusions}
Detailed study of neutrino fluxes from different astrophysical sites such as the
cores of collapsing stars, the galactic plane, as well as other more far away 
anticipated astrophysical and cosmological 
 sites will provide valuable information about the neutrino intrinsic
properties and the site itself.

In collapsing stars, such studies can constrain the role of supernova 
magnetic field in mixed neutrino propagation. 
In particular, if a neutrino spin flavor conversion occurs in the 
isotopically neutral region in a collapsing star, the resulting 
changes in neutrino flux spectrum are sensitive to rather small 
values of neutrino magnetic moment, $\mu $, such as $\mu \leq 10^{-13}\mu_{B}$.
In this context, in the first lecture, after a brief introduction 
of the main neutrino production mechanisms during the supernova stage,
the description for neutrino spin flavor conversions is elaborated
under the assumption of small vacuum mixing. This includes a discussion
that incorporates the possibility of violation of adiabaticity in 
neutrino spin flavor conversions. Results of its implications for
supernova neutrinos are summarized.

For other more distant and energetic 
 sources, 
the prospective high energy neutrino observation will provide clues for a
solution of the long standing problem of origin of observed ultra high energy 
cosmic rays. The (non) observation of high energy neutrinos will also help to
better model the underlying physics of the far away astrophysical and 
cosmological  sites. In this context, in the second lecture, the present
motivations for their searches are presented, with a 
 description of their possible connection 
with ultra high energy cosmic rays. Main high energy neutrino production 
mechanisms are summarized via a simple classification flow chart. Three
flavor neutrino oscillation description is reviewed and its implications
for the three relative ratios of high energy neutrinos are given. Furthermore,
 the essentials of prospective observations of high energy neutrinos are briefly 
described including the possible relevant observational signatures.

The author thanks Physics Division of NCTS for financial support and 
 G.-L. Lin and T.-W. Yeh for comments.

\pagebreak


\begin{thebibliography}{99}
%\tightenlines
%
%
%
\bibitem{Fukugita:wx}
M. Fukugita and T. Yanagida, in {\em Physics And Astrophysics Of
Neutrinos}, ed. by M. Fukugita and A. Suzuki (Springer-Verlag,
Tokyo, 1994), p. 1.
%
%
%
\bibitem{Kim:dy}
C.~W.~Kim and A.~Pevsner, {\em Neutrinos In Physics And Astrophysics}
(Harwood, Chur, Switzerland, 1993) 429 p.
%
%
%
\bibitem{djouadi} F. Boudjema, lectures given in this School;  A. Djouadi, 
 lectures given in this School.
%
%
%
\bibitem{Valle:2002tm}
 See, for instance, J.~W.~Valle, {\em Standard and non-standard neutrino properties},
 invited  talk given at 20th International Conference on Neutrino Physics and 
 Astrophysics (Neutrino 2002), 
 Munich, Germany, 
arXiv:hep-ph/0209047.
%%CITATION = HEP-PH 0209047;%%
%
%
\bibitem{Mikheev:qk}
S.~P.~Mikheev and A.~Yu.~Smirnov,
%``Resonance Oscillations Of Neutrinos In Matter,''
Sov.\ Phys.\ Usp.\  {\bf 30}, 759 (1987) [Usp.\ Fiz.\ Nauk {\bf
153}, 3 (1987)].
%%CITATION = SOPUA,30,759;%%
%
%
%
\bibitem{Wolfenstein:1977ue}
L.~Wolfenstein,
%``Neutrino Oscillations In Matter,''
Phys.\ Rev.\ D {\bf 17}, 2369 (1978);
%%CITATION = PHRVA,D17,2369;%%
S.~P.~Mikheev and A.~Yu.~Smirnov,
%``Resonance Enhancement Of Oscillations In Matter And Solar Neutrino  Spectroscopy,''
Sov.\ J.\ Nucl.\ Phys.\  {\bf 42}, 913 (1985) [Yad.\ Fiz.\  {\bf
42}, 1441 (1985)].
%%CITATION = SJNCA,42,913;%%
%
%
%
%
%
%
\bibitem{Bahcall:gw}
J.~N.~Bahcall, R.~Davis, P.~Parker, A.~Smirnov and
R.~Ulrich, {\em Solar Neutrinos: The First Thirty Years} 
(Addison-Wesley, Reading, 1995), 440 p. 
%
%
%
\bibitem{lipari}
P. Lipari, lectures given in this School;
 G. Giacomelli, lectures given in this School 
(G.~Giacomelli and M.~Sioli, {\em Astroparticle physics},
arXiv:hep-ex/0211035).
%%CITATION = HEP-EX 0211035;%%
%
%
%
\bibitem{Gaisser:2002jj}
T.~K.~Gaisser and M.~Honda,
%``Flux of atmospheric neutrinos,''
arXiv:hep-ph/0203272;
%%CITATION = HEP-PH 0203272;%%
T.~K.~Gaisser, {\em Atmospheric neutrino fluxes},
invited  talk given at 20th International Conference on Neutrino Physics 
 and Astrophysics (Neutrino 2002), 
 Munich, Germany, arXiv:hep-ph/0209195.
%%CITATION = HEP-PH 0209195;%%
%
%
%
\bibitem{Bahcall:iz}
J.~N.~Bahcall and F.~Halzen,
%``Neutrino Astronomy: The Sun And Beyond,''
Phys.\ World {\bf 9}, 41 (1996);
%%CITATION = PHWOE,9N9,41;%%
R.~J.~Protheroe,
%``Developments in high energy neutrino astronomy,''
arXiv:astro-ph/9907374;
%%CITATION = ASTRO-PH 9907374;%%
J.~N.~Bahcall, {\em Astrophysical neutrinos: 20th century and beyond},
IUPAP Centennial lecture given at 19th International Conference on 
 Neutrino Physics and Astrophysics (Neutrino 2000), Sudbury, Canada, 
Nucl.\ Phys.\ Proc.\ Suppl.\  {\bf 91}, 9 (2000)
[arXiv:hep-ph/0009044];
%%CITATION = HEP-PH 0009044;%%
E.~Waxman and K.~Mannheim,
%``High-Energy Neutrinos,''
Europhys.\ News {\bf 32}, 216 (2001);
%%CITATION = EUPNA,32,216;%%
R.~Battiston, {\em Astro particle physics from space},
 invited talk given at ESO - CERN - ESA Symposium on Astronomy, 
Cosmology and Fundamental Physics, 2002, Garching, Germany,  
arXiv:astro-ph/0208108; 
%%CITATION = ASTRO-PH 0208108;%%
H.~Athar, {\em High energy astrophysical neutrinos},
 talk given at IAU 8th Asian - Pacific Regional Meeting, 2002, Tokyo, Japan, 
arXiv:hep-ph/0209130.
%%CITATION = HEP-PH 0209130;%%
%
%
%
\bibitem{Ginzburg:sk}
V.~L.~Ginzburg, V.~A.~Dogiel, V.~S.~Berezinsky, S.~V.~Bulanov and
V.~S.~Ptuskin, {\em Astrophysics Of Cosmic Rays} (North-Holland, 
 Amsterdam, Netherlands, 1990) 534 p.; 
T.~K.~Gaisser, {\em Cosmic Rays And Particle Physics}
 (Cambridge University Press, Cambridge, UK, 1990), 279 p. 
%
%
%
\bibitem{Learned:sw}
J.~G.~Learned and K.~Mannheim,
%``High-Energy Neutrino Astrophysics,''
Ann.\ Rev.\ Nucl.\ Part.\ Sci.\  {\bf 50}, 679 (2000);
%%CITATION = ARNUA,50,679;%%
F.~Halzen and D.~Hooper,
%``High-energy neutrino astronomy: The cosmic ray connection,''
Rept.\ Prog.\ Phys.\  {\bf 65}, 1025 (2002)
[arXiv:astro-ph/0204527].
%%CITATION = ASTRO-PH 0204527;%%
%
%
%
\bibitem{Smirnov:2002in}
A.~Yu.~Smirnov,
 {\em Solar neutrinos: Interpretation of results},
invited  talk given at 20th International Conference on Neutrino 
 Physics and Astrophysics (Neutrino 2002), 
 Munich, Germany, 
arXiv:hep-ph/0209131.
%%CITATION = HEP-PH 0209131;%%
%
%
%
\bibitem{Malek:2002ns}
M.~Malek {\it et al.}  [Super Kamiokande Collaboration],
%``Search for supernova relic neutrinos at Super-Kamiokande,''
arXiv:hep-ex/0209028.
%%CITATION = HEP-EX 0209028;%%
%
%
%
\bibitem{Shapiro:du}
S.~L.~Shapiro and S.~A.~Teukolsky,
 {\em Black Holes, White Dwarfs, And Neutron Stars: The Physics Of Compact Objects}
  (Wiley, New York, 1983), 645 p.
%
%
%
\bibitem{Burrows:kf}
A.~Burrows, D.~Klein and R.~Gandhi,
%``The Future Of Supernova Neutrino Detection,''
Phys.\ Rev.\ D {\bf 45}, 3361 (1992).
%%CITATION = PHRVA,D45,3361;%%
%
%
%
\bibitem{Mohapatra:rq}
R.~N.~Mohapatra and P.~B.~Pal, {\em Massive Neutrinos In Physics And Astrophysics} 
 (Second Edition)
[World Sci.\ Lect.\ Notes Phys.\  {\bf 60}, 1 (1998)].
%%CITATION = 00327,60,1;%%
%
%
%
\bibitem{Raffelt:2002tu}
For a recent discussion, see, G.~G.~Raffelt,
 {\em Physics with supernovae}, invited 
 talk given at Topics in Astroparticle and Underground Physics (TAUP 2001), Assergi, Italy, 
Nucl.\ Phys.\ Proc.\ Suppl.\  {\bf 110}, 254 (2002)
[arXiv:hep-ph/0201099].
%%CITATION = HEP-PH 0201099;%%
%
%
%
\bibitem{Dighe:1999bi}
A.~S.~Dighe and A.~Yu.~Smirnov,
%``Identifying the neutrino mass spectrum from the neutrino burst from a  supernova,''
Phys.\ Rev.\ D {\bf 62}, 033007 (2000) [arXiv:hep-ph/9907423], and
references cited therein.
%%CITATION = HEP-PH 9907423;%%
%
%
%
\bibitem{ahriche}
A. Ahriche and J. Mimouni, poster entitled {\em Probing the neutrino mass spectrum
through Supernova bursts}, presented in this School; S.~Ando and K.~Sato,
%``Three-generation study of neutrino spin-flavor conversion in supernova  and 
%implication for neutrino magnetic moment,''
arXiv:hep-ph/0211053.
%%CITATION = HEP-PH 0211053;%%
%
%
%
\bibitem{Grimus:1993fz}
W.~Grimus and T.~Scharnagl,
%``Neutrino propagation in matter and electromagnetic fields,''
Mod.\ Phys.\ Lett.\ A {\bf 8}, 1943 (1993).
%%CITATION = MPLAE,A8,1943;%%
%
%
%
\bibitem{Notzold:1987ik}
D.~N\"{o}tzold and G. G.~Raffelt,
%``Neutrino Dispersion At Finite Temperature And Density,''
Nucl.\ Phys.\ B {\bf 307}, 924 (1988).
%%CITATION = NUPHA,B307,924;%%
%
%
%
\bibitem{Okun:hi}
M.~B.~Voloshin, M.~I.~Vysotsky and L.~B.~Okun
%``Electromagnetic Properties Of Neutrino And Possible Semiannual Variation Cycle Of 
% The Solar Neutrino Flux,''
Sov.\ J.\ Nucl.\ Phys.\  {\bf 44}, 440 (1986)
[Yad.\ Fiz.\  {\bf 44}, 677 (1986)].
%%CITATION = SJNCA,44,440;%%
%
%
%
\bibitem{Akhmedov:1997yv}
E.~Kh.~Akhmedov,
 {\em The neutrino magnetic moment and time variations of the Solar neutrino  flux},
talk given at 4th International Solar Neutrino Conference, 1997, Heidelberg, Germany, 
arXiv:hep-ph/9705451.
%%CITATION = HEP-PH 9705451;%%
%
%
%
\bibitem{Akhmedov:2002mf}
E.~Kh.~Akhmedov and J.~Pulido,
%``Solar neutrino oscillations and bounds on neutrino magnetic moment and  solar magnetic field,''
arXiv:hep-ph/0209192.
%%CITATION = HEP-PH 0209192;%%
%
%
%
\bibitem{Hagiwara:pw}
K.~Hagiwara {\it et al.}  [Particle Data Group Collaboration],
%``Review Of Particle Physics,''
Phys.\ Rev.\ D {\bf 66}, 010001 (2002).
%%CITATION = PHRVA,D66,010001;%%
%
%
%
\bibitem{Fukugita:1987uy}
M.~Fukugita and S.~Yazaki,
%``Reexamination Of Astrophysical And Cosmological Constraints On The Magnetic 
% Moment Of Neutrinos,''
Phys.\ Rev.\ D {\bf 36}, 3817 (1987);
%%CITATION = PHRVA,D36,3817;%%
 G.~G.~Raffelt,
%``Core Mass At The Helium Flash From Observations And A New Bound On Neutrino 
% Electromagnetic Properties,''
Astrophys.\ J.\  {\bf 365}, 559 (1990);
%%CITATION = ASJOA,365,559;%%
 V.~Castellani and S.~Degl'Innocenti,
%``Stellar Evolution As A Probe Of Neutrino Properties,''
Astrophys.\ J.\  {\bf 402}, 574 (1993).
%%CITATION = ASJOA,402,574;%%
 Other arguments also lead to a similar upper bound, see,
 A.~Ayala, J.~C.~D'Olivo and M.~Torres,
%``Neutrino chirality flip through photon Landau damping in supernovae,''
Phys.\ Rev.\ D {\bf 59}, 111901 (1999)
[arXiv:hep-ph/9804230].
%%CITATION = HEP-PH 9804230;%%
%
%
%
\bibitem{Babu:1991aj}
K.~S.~Babu, R.~N.~Mohapatra and I.~Z.~Rothstein,
%``Reconciling the time dependence of the chlorine and Kamiokande solar neutrino data,''
Phys.\ Rev.\ D {\bf 44}, 2265 (1991).
%%CITATION = PHRVA,D44,2265;%%
%
%
%
\bibitem{LSZ}
The main steps consists in deriving the  $P_{LSZ}$ are to start
from neutrino evolution equation of the type given in Eq. (\ref{spinflipvacuum}) 
for $\nu_{e}$ and $\bar{\nu}_{\mu}$ and
then using appropriate boundary conditions obtain a second order
differential equation for $\bar{\nu}_{\mu}$  (see a  similar consideration 
 for pure flavor case in \cite{Fukugita:wx}). Assuming that the
matter density and magnetic field vary linearly in the resonance region, it is
possible to transform the above differential equation into Weber differential 
equation as follows:
%
%
%
\[
 \frac{{\rm d}^{2}M}{{\rm d}l^{2}}+\left( n+\frac{1}{2}-\frac{l^{2}}{4}\right)M=0,
\]
%
%
%
with $M(r)={\rm Exp}(\frac{\dot{\imath}}{2}
\int \eta r {\rm d}r)c_{2}(r)$ and 
$\bar{\nu}_{\mu}(r)={\rm Exp}(-\dot{\imath}\int \epsilon_{2} {\rm d}r)c_{2}(r)$.
 Here $\epsilon_{2}=\frac{\Delta m^{2}}{4E}$ and $\eta = 
 \sqrt{2}G_{F} \dot{n}^{0}$ with $n(r)=n^{0}+\dot{n}^{0}r$.
 Remaining variables are: $l=\sqrt{\eta}r \, {\rm Exp}(-\dot{\imath}\frac{\pi}{4})$ and 
$n = \dot{\imath}\frac{\epsilon^{2}_{12}}{\eta}$, where $\epsilon_{12}=2\mu B$. 
  Choice of the solution with correct asymptotic behavior 
 of this differential equation gives $\bar{\nu}_{\mu}$
 which in turn can be shown to lead to exponential form of the 
  probability that $\bar{\nu}_{\mu}$ stays as $\bar{\nu}_{\mu}$ after 
 crossing the resonance region, given by Eq. (\ref{lsz}).
%
%
%
\bibitem{Athar:1995cx}
H.~Athar, J.~T.~Peltoniemi and A.~Yu.~Smirnov,
%``Neutrino spin flip effects in collapsing stars,''
Phys.\ Rev.\ D {\bf 51}, 6647 (1995) [arXiv:hep-ph/9501283];
%%CITATION = HEP-PH 9501283;%%
 H.~Athar,
 {\em Spin-flavor conversions of neutrinos in collapsing stars},
talk given at 2nd International Workshop on New Worlds in Astroparticle Physics, 
 1998, Faro, Portugal,
arXiv:hep-ph/9902222, and references cited therein.
%%CITATION = HEP-PH 9902222;%%
%
%
%
\bibitem{Athar:pk}
H.~Athar and J.~T.~Peltoniemi,
%``Signatures Of Neutrino Conversions To Sterile States In Collapsing Stars,''
Phys.\ Rev.\ D {\bf 51}, 5785 (1995).
%%CITATION = PHRVA,D51,5785;%%
%
%
%
\bibitem{Cei:2002mq}
F.~Cei,
%``Neutrinos from supernovae: Experimental status and perspectives,''
Int.\ J.\ Mod.\ Phys.\ A {\bf 17}, 1765 (2002)
[arXiv:hep-ex/0202043].
%%CITATION = HEP-EX 0202043;%%
%
%
%
\bibitem{Athar:2000yw}
H.~Athar, M.~Je\.{z}abek and O.~Yasuda,
%``Effects of neutrino mixing on high-energy cosmic neutrino flux,''
Phys.\ Rev.\ D {\bf 62}, 103007 (2000)
[arXiv:hep-ph/0005104].
%%CITATION = HEP-PH 0005104;%%
%
%
%
\bibitem{Athar:2000ak}
H.~Athar,
%``Neutrino conversions in cosmological gamma ray burst fireballs,''
Astropart.\ Phys.\  {\bf 14}, 217 (2000) [arXiv:hep-ph/0004191].
%%CITATION = HEP-PH 0004191;%%
%
%
%
\bibitem{Keranen:1999nf}
P.~Ker\"{a}nen, J.~Maalampi and J.~T.~Peltoniemi,
%``Testing neutrino instability with active galactic nuclei,''
Phys.\ Lett.\ B {\bf 461}, 230 (1999) [arXiv:hep-ph/9901403];
%%CITATION = HEP-PH 9901403;%%
 J.~F.~Beacom, N.~F.~Bell, D.~Hooper, S.~Pakvasa and T.~J.~Weiler,
%``Decay of high-energy astrophysical neutrinos,''
arXiv:hep-ph/0211305.
%%CITATION = HEP-PH 0211305;%%
%
%
%
\bibitem{Mughal:hr}
M.~A.~Mughal and H.~Athar,
%``Neutrino Spin-Flip Effects In Active Galactic Nuclei,''
Chin.\ J.\ Phys.\  {\bf 40}, 251 (2002).
%%CITATION = CJOPA,40,251;%%
%
%
%
\bibitem{Athar:1999gx}
H.~Athar and J.~F.~Nieves,
%``Matter effects on neutrino oscillations in gravitational and magnetic  fields,''
Phys.\ Rev.\ D {\bf 61}, 103001 (2000) [arXiv:hep-ph/0001069].
%%CITATION = HEP-PH 0001069;%%
%
%
%
\bibitem{Minakata:1996nd}
H.~Minakata and A.~Yu.~Smirnov,
%``High Energy Cosmic Neutrinos and the Equivalence Principle,''
Phys.\ Rev.\ D {\bf 54}, 3698 (1996) [arXiv:hep-ph/9601311].
%%CITATION = HEP-PH 9601311;%%
%
%
%
\bibitem{Weiler:1982qy}
T.~J.~Weiler,
%``Resonant Absorption Of Cosmic Ray Neutrinos By The Relic Neutrino Background,''
Phys.\ Rev.\ Lett.\  {\bf 49}, 234 (1982).
%%CITATION = PRLTA,49,234;%%
%
%
%
\bibitem{Tyler:2000gt}
C.~Tyler, A.~V.~Olinto and G.~Sigl,
%``Cosmic neutrinos and new physics beyond the electro weak scale,''
Phys.\ Rev.\ D {\bf 63}, 055001 (2001)
[arXiv:hep-ph/0002257];
%%CITATION = HEP-PH 0002257;%%
J.~L.~Feng and A.~D.~Shapere,
%``Black hole production by cosmic rays,''
Phys.\ Rev.\ Lett.\  {\bf 88}, 021303 (2002)
[arXiv:hep-ph/0109106].
%%CITATION = HEP-PH 0109106;%%
%
%
%
\bibitem{Berezinsky:kz}
V.~S.~Berezinsky and A.~Yu.~Smirnov,
%``Astrophysical Upper Bounds On Neutrino-Nucleon Cross Section At 
% Energy E>=3 X 10-To-The-17 Ev,''
Phys.\ Lett.\ B {\bf 48}, 269 (1974).
%%CITATION = PHLTA,B48,269;%%
%
%
%
\bibitem{Halzen:1998mb}
F.~Halzen, {\em Lectures on neutrino astronomy: Theory and experiment},
 lectures given at Theoretical Advanced Study Institute in Elementary Particle 
Physics (TASI 98), Boulder, USA, 
arXiv:astro-ph/9810368.
%%CITATION = ASTRO-PH 9810368;%%
%
%
%
\bibitem{Protheroe:1998dm}
R.~J.~Protheroe, {\em High energy neutrino astrophysics},
 invited talk given at 18th International Conference on Neutrino Physics and 
Astrophysics (Neutrino 98), Takayama, Japan,
Nucl.\ Phys.\ Proc.\ Suppl.\  {\bf 77}, 465 (1999)
 [arXiv:astro-ph/9809144].
%%CITATION = ASTRO-PH 9809144;%%
%
%
%
\bibitem{Ferrarese:2002vg}
L.~Ferrarese and D.~Merritt,
%``Super massive black holes,''
arXiv:astro-ph/0206222;
%%CITATION = ASTRO-PH 0206222;%%
Y.~Artemova and I.~Novikov,
%``The Interior of Black Holes and their Astrophysics,''
arXiv:astro-ph/0210545.
%%CITATION = ASTRO-PH 0210545;%%
%
%
%
\bibitem{Waxman:1997ti}
E.~Waxman and J.~N.~Bahcall,
%``High energy neutrinos from cosmological gamma ray burst fireballs,''
Phys.\ Rev.\ Lett.\  {\bf 78}, 2292 (1997)
[arXiv:astro-ph/9701231].
%%CITATION = ASTRO-PH 9701231;%%
%
%
%
\bibitem{Dermer:2000yd}
C.~D.~Dermer,
%``Neutrino, neutron, and cosmic ray production in the external shock  model of gamma ray bursts,''
Astrophys.\ J.\  {\bf 574}, 65 (2002)
[arXiv:astro-ph/0005440].
%%CITATION = ASTRO-PH 0005440;%%
%
%
%
\bibitem{Athar:2002ib}
H.~Athar and G.-L.~Lin, {\em On non hadronic origin of high energy neutrinos},
 invited talk given at 1st NCTS Workshop on Astroparticle Physics, 2001, Kenting, Taiwan, 
arXiv:hep-ph/0203265.
%%CITATION = HEP-PH 0203265;%%
%
%
%
\bibitem{Sigl:2001th}
G.~Sigl, {\em Ultrahigh energy neutrinos and cosmic rays as probes of new physics},
 lectures given at Workshop on Neutrino Physics and Cosmology, 2001, Copenhagen, Denmark, 
arXiv:hep-ph/0109202.
%%CITATION = HEP-PH 0109202;%%
%
%
%
\bibitem{Nagano:ve}
M.~Nagano and A.~A.~Watson,
%``Observations And Implications Of The Ultrahigh-Energy Cosmic Rays,''
Rev.\ Mod.\ Phys.\  {\bf 72}, 689 (2000).
%%CITATION = RMPHA,72,689;%%
%
%
%
\bibitem{Waxman:1998yy}
E.~Waxman and J.~N.~Bahcall,
%``High energy neutrinos from astrophysical sources: An upper bound,''
Phys.\ Rev.\ D {\bf 59}, 023002 (1999)
[arXiv:hep-ph/9807282];
%%CITATION = HEP-PH 9807282;%%
K.~Mannheim, R.~J.~Protheroe and J.~P.~Rachen,
%``On the cosmic ray bound for models of extragalactic neutrino  production,''
Phys.\ Rev.\ D {\bf 63}, 023003 (2001)
[arXiv:astro-ph/9812398];
%%CITATION = ASTRO-PH 9812398;%%
 See, also, 
 O.~E.~Kalashev, V.~A.~Kuzmin, D.~V.~Semikoz and G.~Sigl,
%``Ultra-high energy neutrino fluxes and their constraints,''
Phys.\ Rev.\ D {\bf 66}, 063004 (2002)
[arXiv:hep-ph/0205050].
%%CITATION = HEP-PH 0205050;%%


%
%
%
\bibitem{Greisen:1966jv}
K.~Greisen,
%``End To The Cosmic Ray Spectrum?,''
Phys.\ Rev.\ Lett.\  {\bf 16}, 748 (1966);
%%CITATION = PRLTA,16,748;%%
G.~T.~Zatsepin and V.~A.~Kuzmin,
%``Upper Limit Of The Spectrum Of Cosmic Rays,''
JETP Lett.\  {\bf 4}, 78 (1966)
[Pisma Zh.\ Eksp.\ Teor.\ Fiz.\  {\bf 4}, 114 (1966)].
%%CITATION = JTPLA,4,78;%%
%
%
%
\bibitem{Berezinsky:2000bq}
See, for instance, V.~S.~Berezinsky and V.~I.~Dokuchaev,
%``Hidden source of high-energy neutrinos in collapsing galactic nucleus,''
Astropart.\ Phys.\  {\bf 15}, 87 (2001)
[arXiv:astro-ph/0002274], and references cited therein.
%%CITATION = ASTRO-PH 0002274;%%
%
%
%
\bibitem{Athar:1998ux}
H.~Athar, {\em Tau neutrinos from active galactic nuclei},
 talk given at 5th International Workshop on Tau Lepton Physics (TAU 98), Santander, Spain, 
Nucl.\ Phys.\ Proc.\ Suppl.\  {\bf 76}, 419 (1999)
[arXiv:hep-ph/9811221].
%%CITATION = HEP-PH 9811221;%%
%
%
%
\bibitem{Athar:2002rr}
H.~Athar, {\em High energy astrophysical tau neutrinos: The expectations},
 talk given at 12th International Symposium on Very High Energy Cosmic Ray Interactions 
 (XII ISVHECRI), 2002, 
Geneva, Switzerland,  
arXiv:hep-ph/0210244.
%%CITATION = HEP-PH 0210244;%%
%
%
%
\bibitem{Wichoski:1998kh}
U.~F.~Wichoski, J.~H.~MacGibbon and R.~H.~Brandenberger,
%``High energy neutrinos, photons and cosmic rays from non-scaling cosmic  strings,''
Phys.\ Rev.\ D {\bf 65}, 063005 (2002)
[arXiv:hep-ph/9805419].
%%CITATION = HEP-PH 9805419;%%
%
%
%
\bibitem{Lunardini:2000fy}
C.~Lunardini and A.~Yu.~Smirnov,
%``High-energy neutrino conversion and the lepton asymmetry in the  universe,''
Phys.\ Rev.\ D {\bf 64}, 073006 (2001)
[arXiv:hep-ph/0012056].
%%CITATION = HEP-PH 0012056;%%
%
%
%
\bibitem{Athar:2000tg}
H.~Athar, {\em Cosmic tau neutrinos}, talk given at YITP Workshop on Theoretical Problems
related to Neutrino Oscillations, 2000, Kyoto, Japan, arXiv:hep-ph/0004083.
%%CITATION = HEP-PH 0004083;%%
%
%
%
\bibitem{Maki:mu}
Z.~Maki, M.~Nakagawa and S.~Sakata,
%``Remarks On The Unified Model Of Elementary Particles,''
Prog.\ Theor.\ Phys.\  {\bf 28}, 870 (1962).
%%CITATION = PTPKA,28,870;%%
%
%
%

\bibitem{Athar:1999at}
H.~Athar, {\em High energy cosmic tau neutrinos}, talk given at
 6th International Workshop on Topics in Astroparticle and Underground Physics (TAUP 99), 
 Paris, France,
Nucl.\ Phys.\ Proc.\ Suppl.\  {\bf 87}, 442 (2000)
[arXiv:hep-ph/9912417];
%%CITATION = HEP-PH 9912417;%%
 {\em ibid}, {\em Neutrino conversions in active galactic nuclei},
 talk given at 9th Lomosonov Conference on Elementary Particle Physics, 
 1999, Moscow, Russia, arXiv:hep-ph/0001128.
%%CITATION = HEP-PH 0001128;%%
%
%
%
\bibitem{Athar:2000je}
H.~Athar, {\em Ultra high energy cosmic neutrinos}, talk given at
 8th Asia Pacific Physics Conference (APPC 2000), Taipei, Taiwan, 
arXiv:hep-ph/0008121.
%%CITATION = HEP-PH 0008121;%%
%
%
%
\bibitem{Athar:1999gw}
H.~Athar,
%``Effects of conversions for high energy neutrinos originating from  cosmological 
% gamma-ray burst fireballs,''
arXiv:hep-ph/9901450.
%%CITATION = HEP-PH 9901450;%%
%
%
%
\bibitem{resvanis}
L. Resvanis, {\em High Energy Neutrino Experiments}, 
 talk given at 12th International Symposium on Very
High Energy Cosmic Ray Interactions (XII ISVHERCI), 2002, Geneva,
Switzerland, to appear in its proceedings.
%
%
%
\bibitem{Capelle:1998zz}
K.~S.~Capelle, J.~W.~Cronin, G.~Parente and E.~Zas,
%``On the detection of ultra high energy neutrinos with the Auger  Observatory,''
Astropart.\ Phys.\  {\bf 8}, 321 (1998)
[arXiv:astro-ph/9801313].
%%CITATION = ASTRO-PH 9801313;%%
%
%
%
\bibitem{Chiba:2000fj}
M.~Chiba, T.~Kamijo, M.~Kawaki, H.~Athar, M.~Inuzuka, M.~Ikeda and O.~Yasuda,
 {\em Study of Salt Neutrino Detector}, talk given by M. Chiba at 
1st International Workshop on Radio Detection of High-Energy Particles (RADHEP 2000), 
 Los Angeles, USA, 
AIP Conf.\ Proc.\  {\bf 579}, 204 (2001).
%%CITATION = APCPC,579,204;%%
%
%
%
\bibitem{Athar:2001bi}
H.~Athar and G.-L.~Lin,
%``Implications of anti-nu/e e- $\to$ W- gamma for high-energy anti-nu/e  observation,''
arXiv:hep-ph/0108204 (to appear in Astropart. Phys.);
%%CITATION = HEP-PH 0108204;%%
 {\em ibid}, {\em The implications of $\bar{\nu}_{e} e^{-} \to W^{-} \gamma $ for the 
 detection of  
 high-energy $\bar{\nu}_{e}$}, talk presented by G. -L. Lin 
 at 5th RESCEU International Symposium on New Trends in Theoretical and Observations 
 Cosmology, 
Tokyo, Japan, 2001, 
arXiv:hep-ph/0201026.
%%CITATION = HEP-PH 0201026;%%
%
%
%
\bibitem{Lai:1994bb}
H.~L.~Lai {\it et al.},
%``Global QCD analysis and the CTEQ parton distributions,''
Phys.\ Rev.\ D {\bf 51}, 4763 (1995)
[arXiv:hep-ph/9410404].
%%CITATION = HEP-PH 9410404;%%
For a recent update, see 
J.~Pumplin, D.~R.~Stump, J.~Huston, H.~L.~Lai, P.~Nadolsky and W.~K.~Tung,
%``New generation of parton distributions with uncertainties from global  QCD analysis,''
JHEP {\bf 0207}, 012 (2002)
[arXiv:hep-ph/0201195].
%%CITATION = HEP-PH 0201195;%%
%
%
%
\bibitem{Gandhi:1998ri}
R.~Gandhi, C.~Quigg, M.~H.~Reno and I.~Sarcevic,
%``Ultrahigh-energy neutrino interactions,''
Astropart.\ Phys.\  {\bf 5}, 81 (1996)
[arXiv:hep-ph/9512364], and references cited therein.
%%CITATION = HEP-PH 9512364;%%
%
%
%
\bibitem{Athar:2000rx}
H.~Athar, G.~Parente and E.~Zas,
%``Prospects for observations of high-energy cosmic tau-neutrinos,''
Phys.\ Rev.\ D {\bf 62}, 093010 (2000)
[arXiv:hep-ph/0006123].
%%CITATION = HEP-PH 0006123;%%
%
%
%
\bibitem{Halzen:1998be}
F.~Halzen and D.~Saltzberg,
%``Tau neutrino appearance with a 1000-Megaparsec baseline,''
Phys.\ Rev.\ Lett.\  {\bf 81}, 4305 (1998)
[arXiv:hep-ph/9804354].
%%CITATION = HEP-PH 9804354;%%
%
%
%
\bibitem{Szabo:qx}
A.~P.~Szabo and R.~J.~Protheroe,
%``Implications Of Particle Acceleration In Active Galactic Nuclei For Cosmic Rays And 
% High-Energy Neutrino Astronomy,''
Astropart.\ Phys.\  {\bf 2}, 375 (1994)
[arXiv:astro-ph/9405020].
%%CITATION = ASTRO-PH 9405020;%%
%
%
%
\bibitem{Learned:1994wg}
J.~G.~Learned and S.~Pakvasa,
%``Detecting tau-neutrino oscillations at PeV energies,''
Astropart.\ Phys.\  {\bf 3}, 267 (1995)
[arXiv:hep-ph/9405296].
%%CITATION = HEP-PH 9405296;%%
%
%
%
\bibitem{Feng:2001ue}
J.~L.~Feng, P.~Fisher, F.~Wilczek and T.~M.~Yu,
%``Observability of earth-skimming ultra-high energy neutrinos,''
Phys.\ Rev.\ Lett.\  {\bf 88}, 161102 (2002)
[arXiv:hep-ph/0105067], and references cited therein.
%%CITATION = HEP-PH 0105067;%%
%
%
%
\end{thebibliography}
\end{document}